\documentclass[12pt,preprint]{aastex}
\usepackage{mathptmx}
\usepackage[scaled=0.9]{helvet}
\usepackage{thumbpdf}

\newcommand{\msun}      {\ensuremath{M_\odot}}
\newcommand{\kms}       {\ensuremath{~\mathrm{km~s^{-1}}}}
\newcommand{\au}        {\ensuremath{~\mathrm{AU}}}

\newcommand{\ergs}       {\ensuremath{~\mathrm{erg~s^{-1}}}}
\newcommand{\mhz}       {\ensuremath{~\mathrm{mHz}}}
\newcommand{\hz}       {\ensuremath{~\mathrm{Hz}}}

\newcommand{\units}[1]  {\ensuremath{\mathrm{~{#1}}}}

 \makeatletter

\ifx\undefined\hyperref

\let\url\relax 

\usepackage[pdftitle={Three-Body Dynamics with Gravitational Wave Emission},pdfauthor={Kayhan Gultekin},letterpaper,linktocpage,colorlinks=true,linkcolor={black},pagecolor={black},urlcolor={blue},citecolor={black}]{hyperref}

\else\relax\fi
 \makeatother

\begin{document}

\title{Three-Body Dynamics with Gravitational Wave Emission}
\author{Kayhan G\"{u}ltekin} \author{M. Coleman Miller}
\author{Douglas P. Hamilton} \affil{University of Maryland, College
  Park, Dept.\ of Astronomy}

\begin{abstract}
  We present numerical three-body experiments that include the effects
  of gravitational radiation reaction by using equations of motion
  that include the 2.5-order post-Newtonian force terms, which are the
  leading order terms of energy loss from gravitational waves.  We
  simulate binary-single interactions and show that close approach
  cross sections for three $1~\msun$ objects are unchanged from the
  purely Newtonian dynamics except for close approaches smaller than
  $10^{-5}$ times the initial semimajor axis of the binary. We also
  present cross sections for mergers resulting from gravitational
  radiation during three-body encounters for a range of binary
  semimajor axes and mass ratios including those of interest for
  intermediate-mass black holes (IMBHs).  Building on previous work,
  we simulate sequences of high-mass-ratio three-body encounters that
  include the effects of gravitational radiation.  The simulations
  show that the binaries merge with extremely high eccentricity such
  that when the gravitational waves are detectable by \emph{LISA},
  most of the binaries will have eccentricities $e > 0.9$ though all
  will have circularized by the time they are detectable by LIGO.  We
  also investigate the implications for the formation and growth of
  IMBHs and find that the inclusion of gravitational waves during the
  encounter results in roughly half as many black holes ejected from
  the host cluster for each black hole accreted onto the growing IMBH.

\end{abstract}

\keywords{stellar dynamics --- gravitational waves --- black hole
  physics --- galaxies: star clusters --- globular clusters: general
  --- methods: \emph{n}-body simulations}

\section{Introduction}
\label{intro}
With increasing evidence in support of the existence of
intermediate-mass black holes (IMBHs), interest in these objects as
gravitational wave sources is growing.  With masses $\sim10^{2}$ to
$10^{4}~\msun$, IMBHs are black holes that are more massive than
stellar-mass black holes yet smaller than supermassive black holes
found at the centers of galaxies.  The primary motivation for IMBHs
comes from observations of ultraluminous X-ray sources (ULXs):
extragalactic, non-nuclear, point sources with inferred bolometric
luminosities $L \ga 3\times10^{39}\ergs$ (see \citealt{mc04} for a
review).  Such luminosities are greater than the Eddington luminosity
of a $20~\msun$ object, which is the highest mass black hole that can
be produced with roughly solar metallicity stellar evolution
\citep{fk01}.  The ULXs are thought to be powered by black holes
because many are variable, but they cannot be powered by supermassive
black holes or they would have sunk to the center of their host
galaxies because of dynamical friction.

If, however, the intrinsic luminosity is much smaller than that
inferred from the X-ray flux because the observed emission originates
from a narrow beam directed towards us, then the lower limit on the
mass can fall into stellar-mass black hole range \citep{kingetal01}.
In such a case, these objects would be stellar-mass analogues of
blazars.  There is, however, evidence in support of quasi-isotropic
emission from ULXs.  First, recent observations of a $L_{X} \sim
10^{40}\ergs$ ULX in the Holmberg~II dwarf irregular galaxy show HeII
emission from gas surrounding the ULX and line ratios in agreement
with photoionization from a quasi-isotropically emitting X-ray source,
thus giving weight to the picture that ULXs are more massive than
stellar-mass black holes \citep{kwz04,pm01}.  Second, observations of
ULX spectra that are fit with a combined power-law and multi-colored
disk model indicate disk temperatures much lower than those found in
known stellar-mass black hole X-ray binaries \citep*{mfm04}.  The
temperature should scale as $T \sim M^{-1/4}$, and thus the inferred
temperatures also favor a larger mass.  Finally, \citet{sm03}
discovered a quasi-periodic oscillation (QPO) in the X-ray brightness
of the brightest ULX in M82 (M82~X-1), whose X-ray luminosity is
$L_{X} \approx 8\times10^{40}\ergs$.  The QPO, which has an rms
amplitude of 8.5\% at a centroid frequency of $54\mhz$, is thought to
come from the disk but is too strong to be consistent with a beamed
source that is powered by a stellar-mass black hole.  Thus there is
strong evidence that at least some ULXs cannot be beamed stellar-mass
X-ray binaries.  There has also been theoretical work that suggests
radiation-driven inhomogeneities can allow luminosities up to $10$
times the Eddington limit \citep{begelman02,rb03}, but the most
luminous ULXs would still require a black hole more massive than
stellar-mass black holes to power their bright, variable X-ray
luminosity.

A key to understanding ULXs and IMBHs is their environment.  ULXs are
often found in starburst galaxies and in associations with stellar and
globular clusters.  For example, M82~X-1, one of the most promising
IMBH candidates, is spatially coincident with the young stellar
cluster MGG~11 as determined by near infrared observations
\citep{mgg03}.  Numerical simulations of MGG~11 show that due to its
short dynamical friction timescale compared to the main sequence
lifetime of the most massive stars, runaway growth by collisions
between massive stars can cause a star to grow to $\sim3000~\msun$,
after which it could evolve into an IMBH \citep{pzetal04}.
\citet{fzm01} found a spatial correlation of ULXs with stellar
clusters in the merging Antennae system in excess of that expected
from a uniform distribution of ULXs.  A comparison of Chandra and HST
images of the CD galaxy NGC~1399 at the center of the Fornax cluster
shows a spatial correlation between many of its X-ray point sources
and its globular clusters \citep{alm01}.  These X-ray point sources
include two of the three sources with $L_{X} \ga 2\times10^{39}\ergs$,
and the globular cluster and X-ray positions agree to within the
combined astrometric uncertainties.  In addition, evidence from radial
velocities of individual stars in M15 as well as velocity and velocity
dispersion measurements in G1 indicate that these globular clusters
may harbor large dark masses in their cores \citep[$\sim3000~\msun$
and $1.7~(\pm0.3)\times10^{4}~\msun$,
respectively,][]{gebhardtetal00,grh02,gerssenetal02,grh05}.  For M15
the data cannot rule out the absence of dark mass at the $3\sigma$
level, but the most recent observations of G1 can rule out the absence
of dark mass at the $97\%$ confidence level \citep{grh05}.  In both
cases, the observations are of special interest because they are the
only direct dynamical measurements of the mass of possible IMBHs.
Finally, the Galactic globular cluster NGC~6752 contains two
millisecond pulsars with high, negative spin derivatives in its core
as well as two other millisecond pulsars well into the halo of the
cluster at 3.3 and 1.4 times the half mass radius of the globular
cluster \citep{cmp03,cpg02}.  The pulsars in the cluster core can be
explained by a line-of-sight acceleration by $10^{3}~\msun$ of dark
mass in the central $0.08\units{pc}$ \citep{ferraroetal03a}.  While
the pulsars in the outskirts of the cluster can be explained by
exchange interactions with binary stars, the most likely explanation
is that they were kicked from the core in a close interaction with an
IMBH, either a single IMBH or a binary that contains an IMBH
\citep{cmp03, cpg02}.

One of the most intriguing questions regarding IMBHs is the method of
their formation.  They must form differently than stellar-mass black
holes, which are the result of a core-collapse supernova and have a
maximum mass of 20~\msun, and they are distinct from supermassive
black holes, which have masses $10^{6}~\msun$ to $10^{9}~\msun$ and
are found at the centers of galaxies.  Several studies have found that
IMBHs may form in young stellar clusters where a core collapse leads
to direct collisions of stars \citep{ebisuzakietal01,gfr04,pzm02}.
\citet{mh02a} proposed that IMBHs form from the mergers of
stellar-mass black holes in a dense globular cluster, and
\citet[hereafter Paper~I]{gmh04} expanded this to include the mergers
of stellar-mass black holes with the merger remnant of a young stellar
cluster core collapse.

Of particular interest is the study of IMBHs as sources of detectable
gravitational waves \citep{hpz05,mse04,miller02,will04}.  Orbiting
black holes are exciting candidates for detectable gravitational
waves.  At a distance $d$ a mass $m$ in a Keplerian orbit of size $r$
around a mass $M \gg m$ the gravitational wave amplitude is
\begin{equation}
h \sim \frac{G^{2}}{c^{4}}\frac{Mm}{r d} = 4.7\times10^{-25}
\left(\frac{M}{\msun}\right) \left(\frac{m}{\msun}\right)
\left({\frac{\vphantom{d}r}{\mathrm{AU}}}\right)^{-1}\left({\frac{d}{\mathrm{kpc}}}\right)^{-1},
\label{gwamp}
\end{equation}
where $G$ is the gravitational constant and $c$ is the speed of light.
For comparison, with one-year integrations the \emph{Laser
  Interferometer Space Antenna} (\emph{LISA}) and Advanced LIGO are
expected to reach down to sensitivities of $10^{-23}$ at frequencies
of $10\units{mHz}$ and $100\units{Hz}$, respectively.  Thus binaries
containing IMBHs with $M \ga 100~\msun$ with small separations at
favorable distances are strong individual sources.  During inspiral
the frequency of gravitational waves increases as the orbit shrinks
until it reaches the innermost stable circular orbit (ISCO) where the
orbit plunges nearly radially towards coalescence.  Because of the
quadrupolar nature of gravitational waves, the gravitational wave
frequency for circular binaries is twice the orbital frequency.  At
the ISCO for a non-spinning black hole with $M \gg m$, where $r_{\rm
  ISCO} = 6GM/c^{2}$ and $h \sim Gm/6c^2d$ is independent of the mass
of the primary, the gravitational wave frequency is
\begin{equation}
f_{\rm GW} = 2 f_{\rm orb} = 2\left(\frac{GM}{4\pi^{2}r^{3}_{\rm
      ISCO}}\right)^{1/2} \approx 4400\hz \left(\frac{\msun}{M}\right).
\label{iscofreq}
\end{equation}
Thus a binary with a $100~\msun$ black hole will pass through 
\emph{LISA} band
\citep[$10^{-4}$ to $10^{0}\hz$][]{danzmann00} and into the bands of
ground-based detectors such as LIGO, VIRGO, GEO-600, and TAMA
\citep[$10^{1}$ to
$10^{3}\hz$,][]{fidecaroetal97,schilling98,barish00,andoetal02}
whereas a $1000~\msun$ black hole will be detectable by \emph{LISA}
during inspiral but will not reach high enough frequencies to be
detectable by currently planned ground-based detectors.  After the
final inspiral phase, the gravitational wave signal goes through a
merger phase, in which the horizons cross, and a ringdown phase, in
which the spacetime relaxes to a Kerr spacetime
\citep{fh98a,fh98b,ct02}.
The merger and
ringdown phases emit gravitational waves at a higher frequency with a
characteristic ringdown frequency of $f\sim10^{4}~\hz (M/\msun)^{-1}$
so that mergers with more massive IMBHs will still be detectable with
ground-based detectors.

IMBHs in dense stellar systems are a unique source of gravitational
waves.  Through mass segregation, the most massive objects will sink
to the center of a stellar cluster as dynamical interactions between
objects tend towards equipartion of energy.  For a cluster that is old
enough to contain compact stellar remnants, the stellar-mass black
holes, including those in binary systems, and IMBHs will congregate at
the center and interact more frequently.  For any reasonable mass
function, the total mass of black holes in a cluster is large enough
that mass segregation is a runaway process, known as the ``mass
stratification instability'' \citep[e.g.,][]{olearyetal05}.  Through
an exchange bias in which the most massive objects tend to end up with
a companion after a three-body encounter, IMBHs will swap into
binaries.  Thus in a dense stellar system with an IMBH and massive
binaries, the IMBH is likely to be found in the binary and to be a
potential strong source of gravitational waves.  The IMBH-containing
binary will continue to interact with objects in the cluster and to
acquire ever more massive companions.  In Paper~I we found that a
binary with an IMBH that undergoes repeated interactions in a stellar
cluster will have a very high eccentricity after its last encounter
before merging, and a significant fraction will retain a measurable
eccentricity ($0.1 \la e \la 0.2$) when they are most easily
detectable with \emph{LISA} \citep[see also][]{olearyetal05}.  Because
detection of inspiral requires the comparison of the signal to a
pre-computed waveform template that depends on the orbital properties
of the binary, knowing the eccentricity distribution is useful.  For
$e \la 0.2$, circular templates are accurate enough to detect the
gravitational wave signal with LIGO \citep{mp99}, and this is likely
to be the case for \emph{LISA} as well.

If stellar clusters frequently host IMBHs, then currently planned
gravitational wave detectors may detect mergers within a reasonable
amount of time.  Optimistic estimates put the upper limit to the
Advanced LIGO detection rate of all black holes in dense stellar
clusters at $\sim10\units{yr^{-1}}$ \citep{olearyetal05}.  The
\emph{LISA} detection rate for $1\units{yr}$ integration and
signal-to-noise ratio of ${\rm S/N} = 10$ is \citep{will04}
\begin{equation}
  \nu_{\rm det} \approx
  10^{-6} \left(\frac{H_{0}}{70\units{km~s^{-1} Mpc^{-1}}}\right)^{3} \left(\frac{f_{\rm
        tot}}{0.1}\right) \left(\frac{\mu}{10~\msun}\right)^{19/8}
  \left(\frac{M_{\rm max}}{100~\msun}\right)^{13/4}
  \left(\ln{\frac{M_{\rm max}}{M_{\rm min}}}\right)^{-1}\units{yr^{-1}},
\end{equation}
where $H_{0}$ is the Hubble constant, $f_{\rm tot}$ is the total
fraction of globular clusters that contain IMBHs, $\mu$ is the reduced
mass of the merging binary, and $M_{\rm min}$ and $M_{\rm max}$ denote
the range in masses of IMBHs in clusters.  If we assume that $M_{\rm
  max} = 10^{3}~\msun$, $\mu = 10~\msun$, $f_{\rm tot} = 0.8$
\citep{olearyetal05}, $H_{0} = 70\units{km~s^{-1} Mpc^{-1}}$, and
$M_{\rm min} = 10^{2}~\msun$, then we get a rate of
$0.006\units{yr^{-1}}$.  This, however, implies that $10^{3}~\msun$
black holes are continuously accreting $10~\msun$ black holes, which
is unlikely to be the case.  Since the distance out to which
\emph{LISA} can detect a given gravitational wave luminosity $D_{L}$
scales as the square root of the integration time $T$, the volume
probed scales as $V \sim D_{L}^{3} \sim T^{3/2}$.  This means that a
$10\units{yr}$ integration could yield a rate of $0.2\units{yr^{-1}}$,
and if IMBHs with mass $M = 10^{4}~\msun$ are common, the rate could
be much higher.  These rates, however, are optimistic and should be
considered as upper limits.  A gravitational wave detection of an IMBH
with high signal-to-noise could also yield the spin parameter and thus
shed light on the formation mechanism of the IMBH \citep{miller02}.  A
full understanding of the gravitational wave signals from IMBHs
requires a more detailed study of the complicated dynamics and
gravitational radiation of these systems.

In this paper we present a study of the dynamics of black holes in a
stellar cluster using numerical simulations that include the effects
of gravitational radiation.  We include gravitational radiation
reaction by adding a drag force to the Newtonian gravitational
calculation.  Our treatment is similar to that of \citet{lee93}, but
we focus on individual encounters and sequences of encounters and the
resulting mergers instead of ensemble properties of the host cluster.
Paper~I incorporated gravitational radiation by integrating the
\citet{peters64} orbit-averaged equations for orbital evolution of a
binary that is emitting gravitational waves, but in this paper we
include the energy loss from gravitational radiation for arbitrary
motion of the masses.  Although the vast majority of three-body
interactions do not differ greatly from a purely Newtonian simulation,
an important few involve close approaches in which gravitational waves
carry away a dynamically significant amount of energy such that it may
cause the black holes to merge quickly in the middle of the encounter.
This is qualitatively different from the mergers in Paper~I which were
caused by gravitational waves emitted by isolated binaries between
encounters, and this new effect is important in considering detectable
gravitational waves as well as IMBH growth.

In \S~\ref{nummeth} we describe our method of including gravitational
waves as a drag force as well as numerical tests of its accuracy.  We
present our simulations and major results in \S~\ref{simulations} and discuss 
the implications for IMBH formation and gravitational wave detection 
in \S~\ref{discussion}.

\section{Numerical Method}
\label{nummeth}

\begin{figure}
\epsscale{0.75}
\rotatebox{90}{\plotone{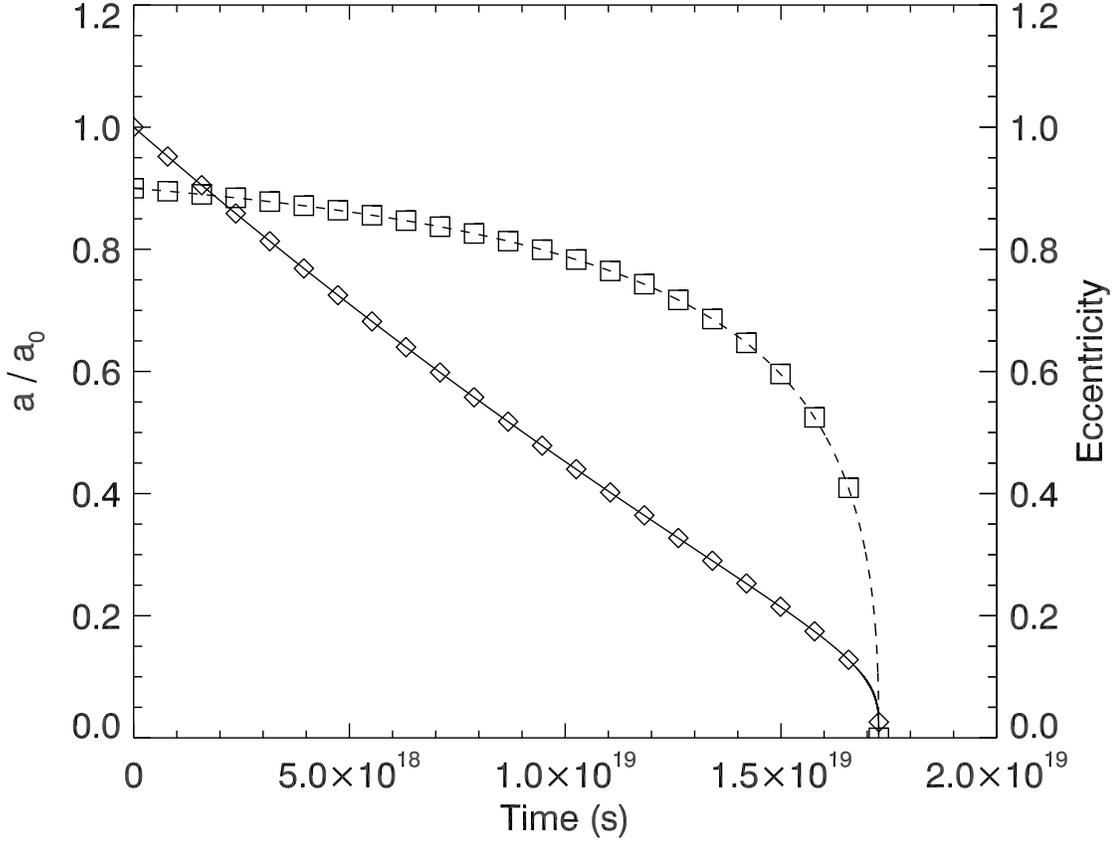}}
\caption{Comparison of HNDrag integration with numerical integration
  of \protect{\citet{peters64}} equations for an eccentric binary.
  Lines are numerical integration of Eq.~\ref{petersa} for semimajor
  axis (solid line) and of Eq.~\ref{peterse} for eccentricity (dashed
  line).  The symbols are results from HNDrag integration with
  gravitational radiation for semimajor axis (diamonds) and
  eccentricity (squares).  The binary shown has $m_{0} = m_{1} =
  10~\msun$ with an initial orbit of $a_{0} = 1\au$ and $e_{0} = 0.9$.
  The evolution of the binary's orbital elements is in very close
  agreement for the entire life of the binary.}
\label{peterscompecc}
\end{figure}

The numerical method we use here is much the same as is described in
\S~2 of Paper~I.  In order to study the dynamics of a massive binary
in a dense stellar environment, we simulate the encounters between the
binary and single objects.  We include both individual encounters and
sequences of encounters, all of which include gravitational radiation
emission.  When simulating sequences, we allow the properties of the
binary to evolve from interactions with singles, and we follow the
binary until a merger occurs.  A merger is determined to occur when
the separation between the two masses is less than $G (m_{0} + m_{1})
/ c^{2}$.  The simulations are run using the same code as in Paper~I
with a few exceptions.  The integration engine is now HNDrag, which is
an extension of HNBody (K.~Rauch \& D.~Hamilton, in
preparation)\footnote{See http://janus.astro.umd.edu/HNBody/.}.  Both
HNBody and HNDrag can include the first-order post-Newtonian
corrections responsible for pericenter precession based on the method
of \citet{nsw83}.  HNDrag also has the ability to include pluggable
modules that can add extra forces or perform separate calculations
such as finding the minimum separation between all pairs of objects.
In this paper we ignore the second-order post-Newtonian terms, which
contribute higher-order corrections to the pericenter precession, and
we include the effects of gravitational radiation on the dynamics of
the particles through the addition of a force that arises from the
$2.5$-order post-Newtonian equation of motion for two point masses.
The acceleration on a mass $m_{0}$ from gravitational waves emitted in
orbit around a mass $m_{1}$ can be written as
\begin{equation}
\label{ddgwdrag}
\frac{d{\mathbf v_{0}}}{d\tau} = 
\frac{4 G^{2}}{5 c^{5}} \frac{m_{0} m_{1}}{r^3}
\left[{\hat{r}}\left({\hat{r}}\cdot{\mathbf v}\right) \left( -6
\frac{Gm_{0}}{r} + \frac{52}{3}\frac{Gm_{1}}{r} + 3v^2\right)
+ {\mathbf v} \left(2 \frac{Gm_{0}}{r} - 8\frac{Gm_{1}}{r}
-v^2\right)\right]
\end{equation}
where ${\mathbf r} = {\mathbf r_{1}} - {\mathbf r_{0}}$ and ${\mathbf
  v} = {\mathbf v_{1}} - {\mathbf v_{0}}$ are the relative position
and velocity vectors between the two masses \citep[][for more recent
treatments that use different techniques and arrive at the same
result, see \citealt*{ifa01} and
\citealt*{bfp98}]{dd81,damour82,damour83}.  Since
Equation~\ref{ddgwdrag} introduces a momentum flux on the center of
mass of the system, we partition the force so that it is equal and
opposite and for reasons of computational efficiency to
get an acceleration of
\begin{equation}
\label{hndgwdrag}
\frac{d{\mathbf v_{0}}}{dt} = 
\frac{4 G^{2}}{5 c^{5}} \frac{m_{0} m_{1}}{r^3} 
\left(\frac{m_{1}}{m_{0} + m_{1}}\right)
\left[\hat{r} \left(\hat{r}\cdot{\mathbf v}\right)
\left(\frac{34}{3} \frac{G\left(m_{0}+m_{1}\right)}{r} + 6v^{2} \right)
+ {\mathbf v} \left( -6\frac{G\left(m_{0}+m_{1}\right)}{r} - 2v^{2}\right)\right].
\end{equation}
This expression is equivalent to Eq.~21 from \citet{lee93}.  When
orbit-averaged, Equation~\ref{hndgwdrag} gives the \citet{peters64}
equations for semimajor axis and eccentricity evolution:
\begin{equation}
  \frac{da}{dt} = - \frac{64}{5}
     \frac{G^{3} m_{0} m_{1} \left(m_{0} + m_{1}\right)}{c^{5} a^{3}
       \left(1 - e^{2}\right)^{7/2}}
     \left( 1 + \frac{73}{24}e^{2} + \frac{37}{96}e^{4}\right)
  \label{petersa}
\end{equation}
\begin{equation}
  \frac{de}{dt} = - \frac{304}{15}
     \frac{G^{3} m_{0} m_{1} \left(m_{0} + m_{1}\right)}{c^{5} a^{4}
       \left(1 - e^{2}\right)^{5/2}} \left(e + \frac{121}{304}e^{3}\right).
  \label{peterse}
\end{equation}
We tested the inclusion of this force in the integrator by comparison
with direct, numerical integration of Equations~\ref{petersa} and
\ref{peterse} for two different binaries with masses $m_{0} = m_{1} =
10~\msun$ and initial semimajor axis $a_{0} = 1\au$: one with initial
eccentricity $e_{0} = 0$ and one with initial eccentricity $e_{0} =
0.9$ (Fig.~\ref{peterscompecc}).  The \emph{N}-body integration of
these binaries made use of HNDrag's enhancement factor, which
artificially augments the magnitude of the drag forces for the
purposes of testing or simulating long-term effects.  For this test
and all numerical integrations with HNDrag, we used the fourth-order
Runge-Kutta integrator.  For both the
circular and the high eccentricity cases, the \emph{N}-body
integrations agree very well with the \cite{peters64} equations.
Examination of Equation~\ref{hndgwdrag} reveals that even though
physically the emission of gravitational radiation can only remove
energy from the system, the equation implies $\dot{E} > 0$ for
$\hat{r}\cdot\hat{v} > 0$ in hyperbolic orbits, becoming worse as the
eccentricity increases \citep{lee93}.  Integration of
Equation~\ref{hndgwdrag} over an entire orbit, however, does lead to
the expected energy loss.  This is because there is an excess of
energy loss at pericenter, which cancels the energy added to the
system \citep{lee93}.  Thus this formulation does not introduce
significant error as long as the integration is calculated accurately
at pericenter, which we achieve by setting HNDrag's relative accuracy
parameter to $10^{-13}$, and the two objects are relatively isolated,
which we discuss below.

\begin{figure}
\epsscale{0.75}
\rotatebox{90}{\plotone{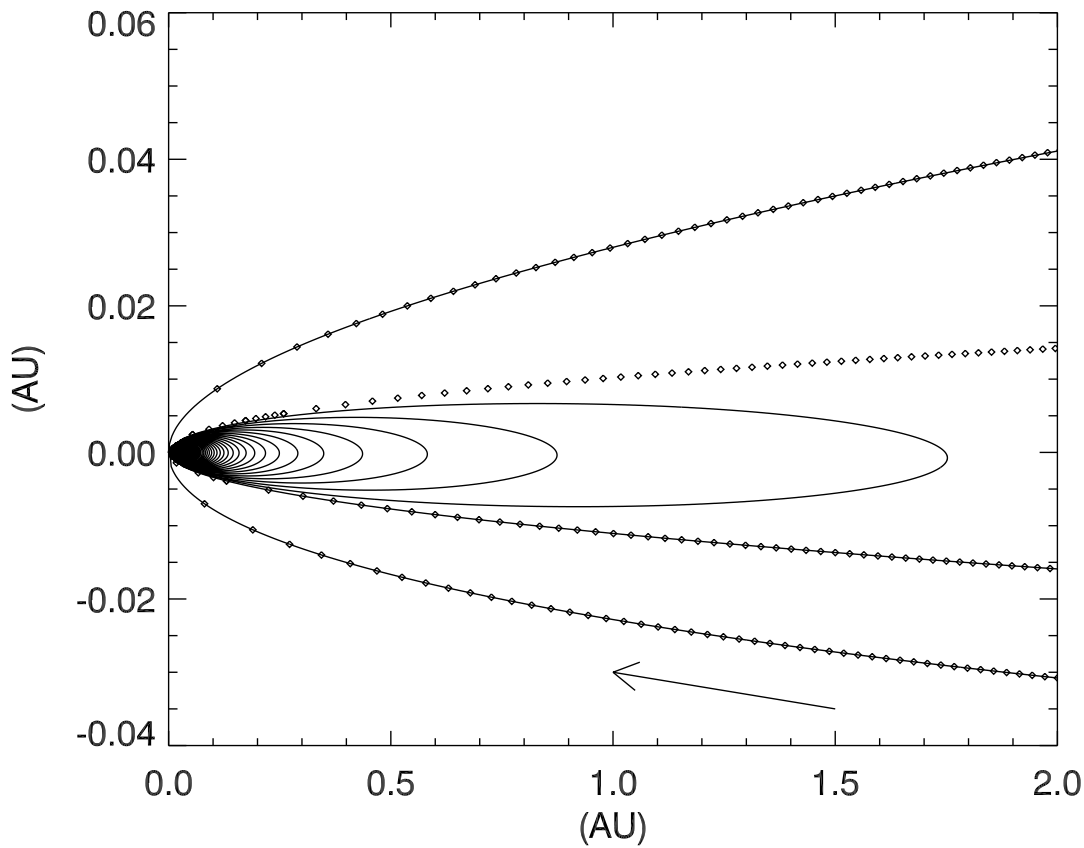}}
\caption{HNDrag-integrated orbits with and without gravitational
  radiation inside and outside of two-body capture pericenter.  This
  plot shows orbits of two $10~\msun$ black holes with relative
  velocity of $10\kms$ and pericenter distances of $r_{p} = 0.2 r_{p,
    {\rm max}} $ and $r_{p} = 1.2 r_{p, {\rm max}}$. The lines show
  the orbits with gravitational radiation included in the integration,
  and the diamonds show the Newtonian orbits for the same initial
  conditions.  The direction of the orbit is indicated by the arrow.
  Although it is not apparent for the outer orbit in this plot, both
  trajectories differ from their Newtonian counterparts.  For the
  inner orbit, enough energy is radiated away for the black holes to
  become bound to each other and eventually merge.}
\label{qscomp}
\end{figure}

We also tested the \emph{N}-body integration with gravitational radiation
for unbound orbits against the maximum periastron separation for two
objects in an initially unbound orbit to become bound to each other
\citep{qs89}:
\begin{equation}
\label{maxpericapture}
r_{p, {\rm max}} = \left( \frac{85\pi\sqrt{2}G^{7/2}m_{0}m_{1}
\left(m_{0} + m_{1}\right)^{3/2}}
 {12c^{5}v^{2}_{\infty}}\right)^{2/7},
\end{equation}
where $v_{\infty}$ is the relative velocity at infinity of the two
masses.  In Figure~\ref{qscomp} we plot the orbits integrated both
with and without gravitational radiation for two different sets of
initial conditions that straddle the $r_{p, {\rm max}}$ threshold.
For both sets of initial conditions, the integrations with
gravitational radiation differ from the Newtonian orbits, and the
inner orbit loses enough energy to become bound and ultimately merge.
We used a bisection method of multiple integrations to calculate
$r_{p, {\rm max}}$, and our value agrees with that of \citet{qs89} to
a fractional accuracy of better than $10^{-5}$.

For systems of three or more masses, we compute gravitational
radiation forces for each pair of objects and add them linearly.
Although this method differs from the full relativistic treatment,
which is nonlinear, the force from the closest pair almost
always dominates.  We may estimate the probability of a third object
coming within the same distance by examining the timescales for an
example system. A binary black hole system with $m_{0} = 1000~\msun$
and $m_{1} = 10~\msun$ with a separation $a = 10^{-2}\au$ ($\sim
1000M$) will merge within \citep{peters64}
\begin{equation}
  \tau_{\rm merge} \approx 6\times10^{17}
  \frac{\left(1~\msun\right)^{3}} {m_{0} m_{1} \left(m_{0} + m_{1}\right)} 
  \left(\frac{a}{1~\mathrm{AU}}\right)^{4} 
  \left(1 - e^{2}\right)^{7/2} \mathrm{yr} \approx 600~\mathrm{yr}.
  \label{mergertime}
\end{equation}  
The expression for merger time in Equation~\ref{mergertime} is valid
for the high eccentricities ($e \rightarrow 1$) of interest to this paper.
The rate of gravitationally focused encounters with a third mass
$m_{2}$ within a distance $r$ from an isotropic distribution is
(Paper~I)
\begin{equation}
  {\nu_{\rm enc}} = 5 \times 10^{-8} 
    \left(\frac{10~\mathrm{km\;s^{-1}}}{v_{\mathrm{\infty}}}\right) 
    \left(\frac{n}{10^{6}~\mathrm{pc^{-3}}}\right) 
    \left(\frac{r}{1~\mathrm{AU}}\right) 
    \left(\frac{m_{0} + m_{1}}{1~\msun}\right)
    \left(\frac{m_{2}}{1~\msun}\right)^{1/2} \mathrm{yr^{-1}}.
  \label{enctime}
\end{equation}
For a number density $n = 10^{6}~{\rm pc^{-3}}$, a relative velocity
$v_{\infty} = 10\kms$, and an interloper mass $m_{2} = 10~\msun$, the
rate of encounters within the same distance $r = a = 10^{-2}\au$ is
${\nu_{\rm enc}} \sim 2\times10^{-6}~{\rm yr^{-1}}$.  Thus the
probability of an encounter within the same distance is
$P\sim\tau_{\rm merge}{\nu_{\rm enc}} \approx 10^{-3}$ for this mildly
relativistic case.  For a separation of $10^{-3}\au$, the probability
drops to $10^{-8}$.  Thus for most astrophysical scenarios and for all
simulations in this paper, the error incurred from adding the
gravitational radiation force terms linearly is negligible.

\section{Simulations and Results}
\label{simulations}

\subsection{Individual Binary-Single Encounters}
\label{individual}

\subsubsection{Close Approach}
\label{closeapproach}

\begin{figure}
\epsscale{0.75}
\rotatebox{90}{\plotone{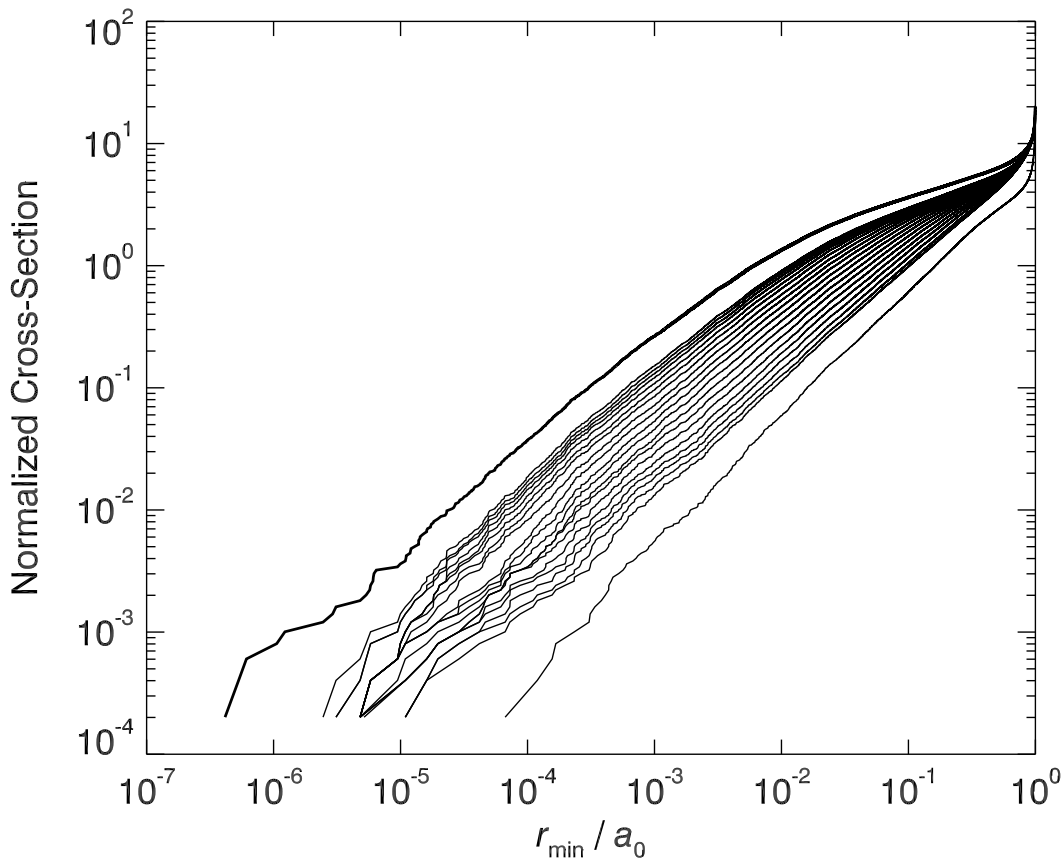}}
\caption{Cross section for close approach during binary-single
  encounters as a function of $r_{\rm min} / a_{0}$.  The thick, upper
  curve is the cross section for the entire encounter.  The remaining
  curves are the cross section at intermediate, equally-spaced times
  during the encounter starting from the bottom near the time of
  initial close approach.  Because we only include 20 intermediate
  curves, there is a gap between the last intermediate curve and the
  final curve.  The first two curves have a slope of $1$, consistent
  with close approach dominated by gravitational focusing.  As the
  encounters progress, resonant encounters with multiple passes are
  more likely to have a close approach at smaller $r_{\rm min} /
  a_{0}$, and the curves gradually evolve to the total cross section
  for the entire encounter. }
\label{newtclose}
\end{figure}

\begin{figure}
\epsscale{0.75}
\rotatebox{90}{\plotone{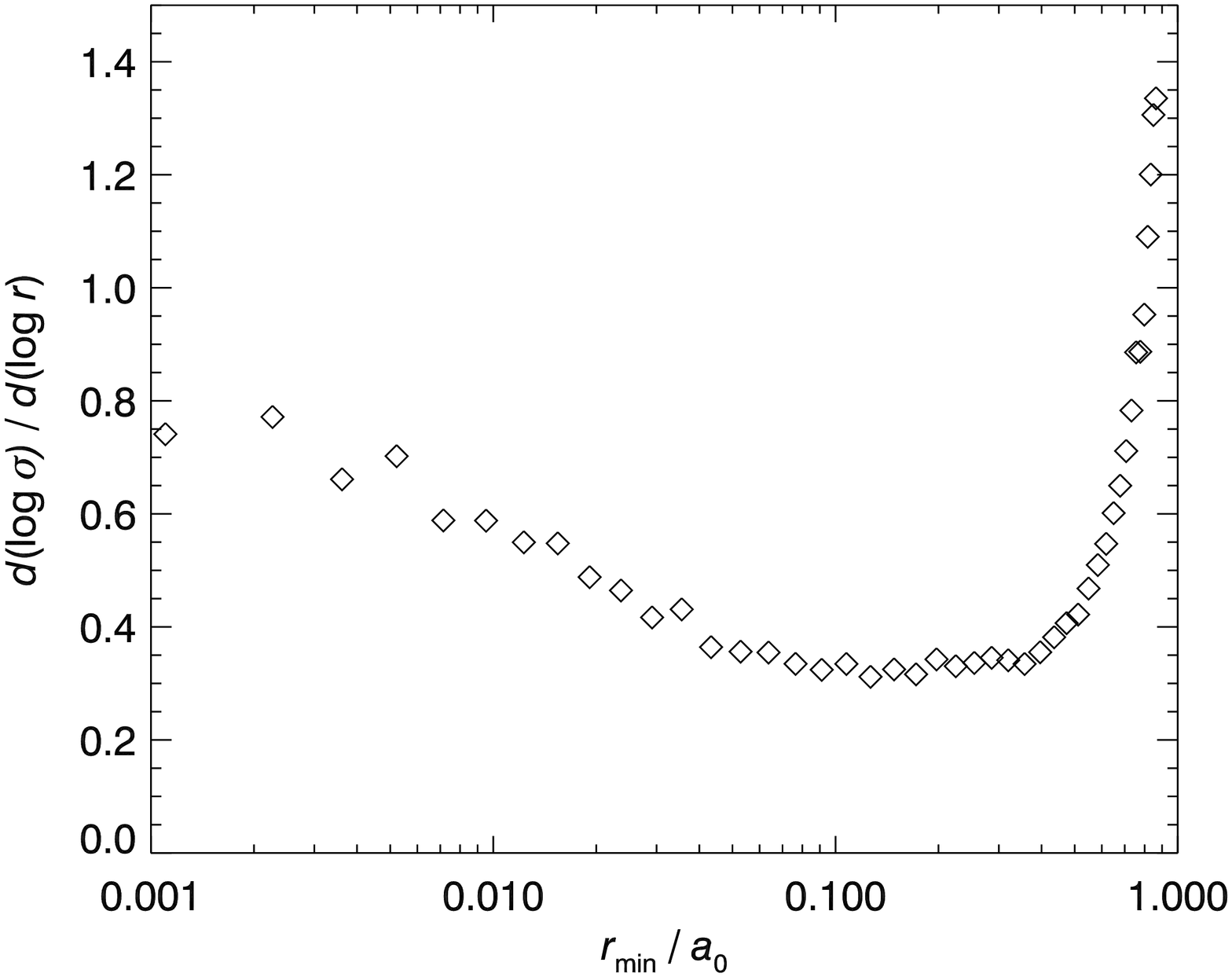}}
\caption{Derivative of close approach cross section curve for the
  entire encounter.  Each symbol is the slope for a line segment fit
  to the top curve from Fig.~\ref{newtclose} plotted as a function of
  the midpoint of the range.  Because of the small number of
  encounters that result in very small close approaches, the multiple
  line segments used in the fits cover different ranges in
  $\log{\left(r/a_{0}\right)}$.  They were selected so that each of
  the 100 line segments covers an additional 1000 encounters that make
  up the cumulative cross section curve.  The scatter in the points is
  indicative of the statistical uncertainty.  For smaller close
  approaches, $d\left(\log{\sigma}\right) / d\left(\log{r}\right)$
  appears to approach unity.  The rise at the right occurs because the
  cross section is formally infinite at $r_{\rm min} / a_{0} = 1$.  }
\label{closeslopes}
\end{figure}

We begin our study of three-body encounters including gravitational
radiation by calculating the minimum distance between any two objects
during the binary-single scattering event.  This quantity has been
well studied for the Newtonian case, but it is still not completely
understood \citep{hi85,sp93}.  We present $10^{5}$ simulations of a
circular binary with masses $m_{0} = m_{1} = 1~\msun$ and an initial
semimajor axis $a_{0} = 1\au$ interacting with an interloper of mass
$m_{2} = 1~\msun$ in a hyperbolic orbit with respect to the center of
mass of the binary.  Throughout this paper, we refer to the mass
ratios of three-body encounters as $m_{0}$:$m_{1}$:$m_{2}$, where
$m_{2}$ is the interloper and the binary consists of $m_{0}$ and
$m_{1}$ with $m_{\rm bin} = m_{0} + m_{1}$ and $m_{0} \ge m_{1}$.  The
relative velocity of the binary and the interloper at infinity is
$v_{\infty} = 0.5\kms$ with an impact parameter randomly drawn from a
distribution with a probability $P(b)\propto b$ and a maximum value
$b_{\rm max} = 6.621\au$, which corresponds to a two-body pericenter
distance of $r_{p} = 5a$.  The encounters are integrated until
finished as determined in Paper~I while tracking the minimum distances
between all pairs of objects.  We follow \citet{hi85} and \citet{sp93}
in calculating a cumulative, normalized cross section for close
approach less than $r$
\begin{equation}
\label{normsigma}
\sigma\left(r\right) = \frac{f\left(r\right)b_{\rm max}^{2}}
{a_{0}^{2}}\left(\frac{v_{\infty}}{v_{c}}\right)^{2},
\end{equation}
where 
\begin{equation}
\label{criticalveloc}
v_{c} \equiv \sqrt{G\frac{m_{0}m_{1}}{a m_{2}}
 \frac{\left(m_{0}+m_{1}+m_{2}\right)}{\left(m_{0}+m_{1}\right)}}
\end{equation}
is the minimum relative velocity required to ionize the system and
$f(r)$ is the fraction of encounters that contain a close approach
less than $r$.  We plot $\sigma\left(r/a_{0}\right)$ for the Newtonian
case at several different time intervals within the encounter in
Figure~\ref{newtclose}.  Our results for the total cross section are
in almost exact agreement with \citet{sp93} over the domain of
overlap, but with the advantage of ten years of computing advances, we
were able to probe down to values of $r/a_{0}$ that are $10^{2}$ times
smaller. In addition we examine how the total cross section evolves
from the initial close approach of the binary until the end of the
interaction through subsequent near passes during long-lived resonant
encounters.  At the time of the interloper's initial close approach
with the binary, the cross section is dominated by gravitational
focusing, and thus the first two curves in Figure~\ref{newtclose} are
well fit by power laws with slope of $1$.  As the interactions
continue, resonant encounters with multiple close approaches are
possible, and the cross section for small values of $r/a_{0}$
increases.  Each successive, intermediate curve approaches the final
cross section by a smaller amount because there are fewer encounters
that last into the next time bin.  A fit of two contiguous power laws
to the final curve yields a break at $r/a_{0} = 0.0102$ with slopes of
$0.85$ and $0.35$ for the lower and upper portions, respectively.
These values are very close to those obtained by earlier studies
\citep{hi85,sp93}.  There is, however, no reason for a preferred scale
for a Newtonian system, and simple models that assume close approaches
are dominated by pericenter passage after an eccentricity kick cannot
explain the lower slope.  We numerically calculate
$d\left(\log{\sigma}\right) / d\left(\log{r}\right)$ by fitting
multiple lines to $\sigma\left(r\right)$ in logarithmic space and plot
the results in Figure~\ref{closeslopes}.  The derivative
$d\left(\log{\sigma}\right) / d\left(\log{r}\right)$ appears to
approach unity for very small values of $r/a_{0}$ where the close
approach can be thought of as a gravitationally focused two-body
encounter within the entire system \citep{hi85}.  It is surprising
that this does not happen until $r/a_{0} < 10^{-5}$.

\begin{figure}
\epsscale{0.75}
\rotatebox{90}{\plotone{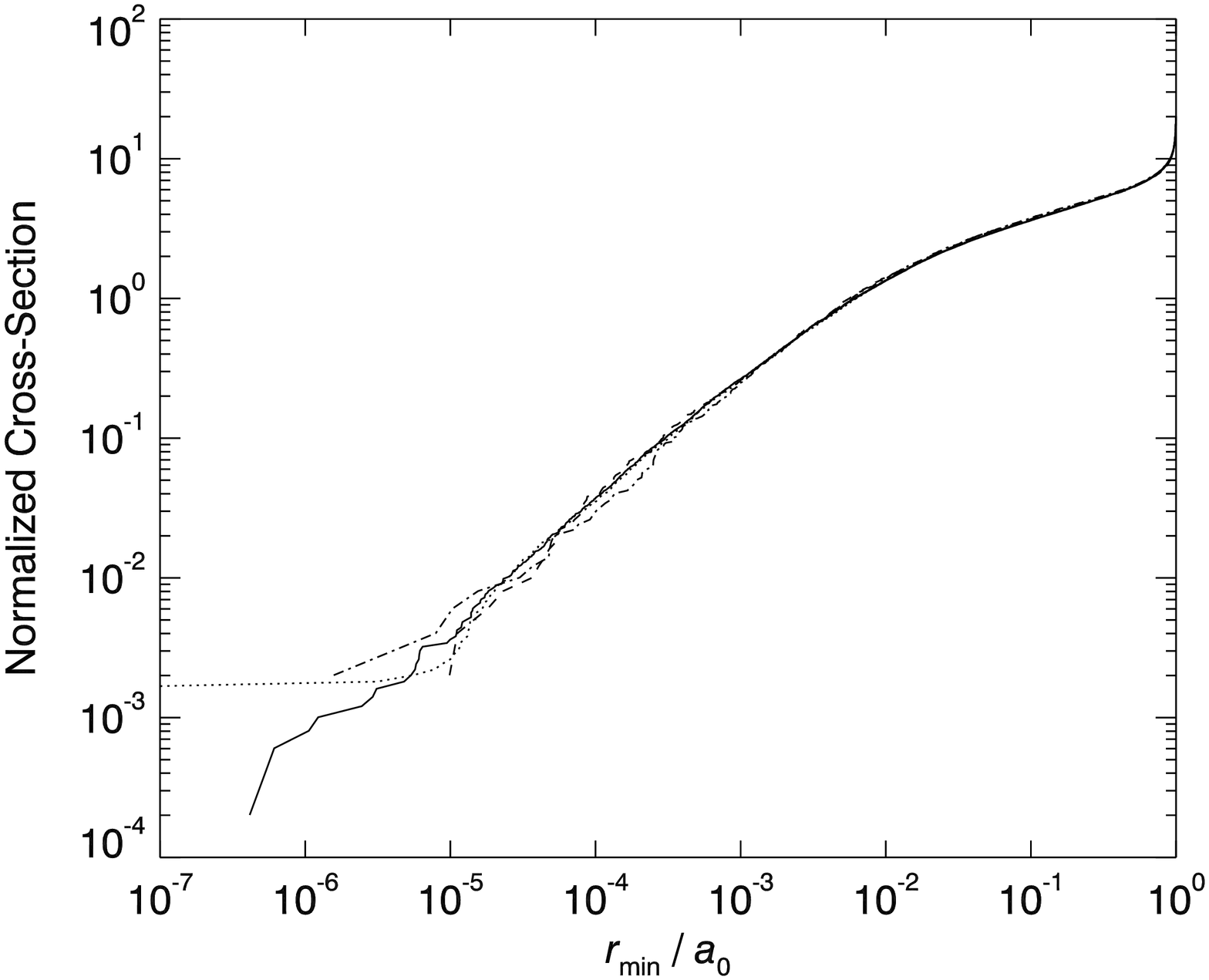}}
\caption{Cross section for close approach like Fig.~\ref{newtclose}
  including different orders of Post-Newtonian corrections.  The
  curves are purely Newtonian (solid), Newtonian plus 2.5-order PN
  (dotted), Newtonian plus 1-order PN (dashed), and Newtonian plus
  1-order and 2.5-order PN (dash-dotted).  The purely Newtonian and
  the Newtonian plus 2.5-order PN curves come from $10^{5}$ encounters
  each.  The other two curves come from $10^{4}$ encounters each and
  show more statistical fluctuations.  The differences between the
  curves are not statistically significant.  }
\label{grclose}
\end{figure}

In order to test the effects of gravitational radiation on the close
approach as well as to test the sensitivity of the results to the
phase of the binary, we ran the same simulations (1) with
gravitational radiation, (2) with gravitational radiation and
first-order post-Newtonian corrections, and (3) with just first-order
post-Newtonian corrections.  The three new cross sections
are plotted with the Newtonian results in Figure~\ref{grclose}.  A K-S
test shows the differences between the three curves to be
statistically insignificant ($P\ge0.4$).  Although not statistically
significant, the curves with gravitational radiation appear to drop
below the Newtonian curve for small $r/a_{0}$ and then climb above for
very small $r/a_{0}$.  Gravitational radiation causes this effect
by driving objects that become very close to each other closer still
and, in some cases, causing them to merge.  For larger masses, the
gravitational radiation is stronger, and the gravitational radiation
curve will differ from the Newtonian curve at larger $r/a_{0}$ for a
fixed value of $a_{0}$.

\subsubsection{Merger Cross Section}
\begin{figure}
\epsscale{0.75}
\rotatebox{90}{\plotone{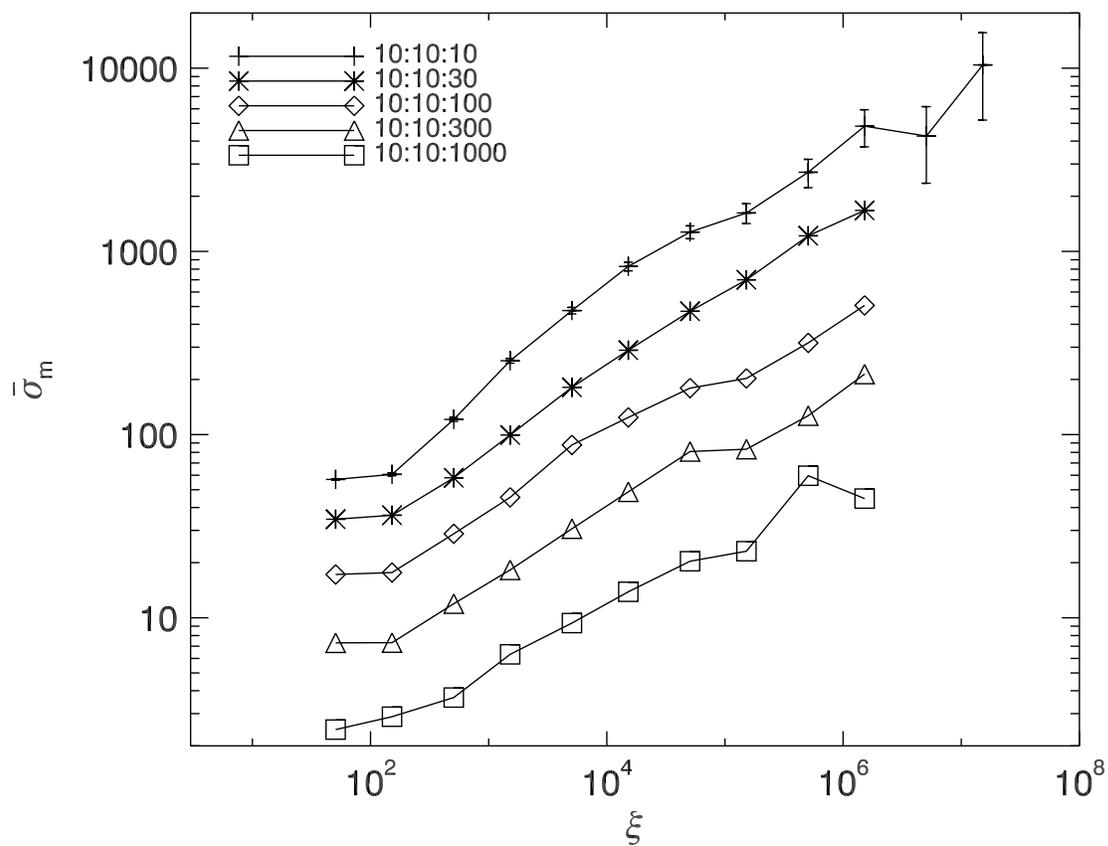}}
\caption{Normalized merger cross sections (Eq.~\ref{sigmabar}) for individual binary-single
  encounters as a function of $\xi$ for the 10:10:X mass series.  The
  normalization is explained in the text.  Each symbol represents
  $10^{4}$ binary-single encounters.  Error bars given for the top
  curve are representative for all merger cross section curves in
  Figures~\ref{mergercs1010x} through \ref{mergercs1000x1000}.}
\label{mergercs1010x}
\end{figure}

\begin{figure}
\epsscale{0.75}
\rotatebox{90}{\plotone{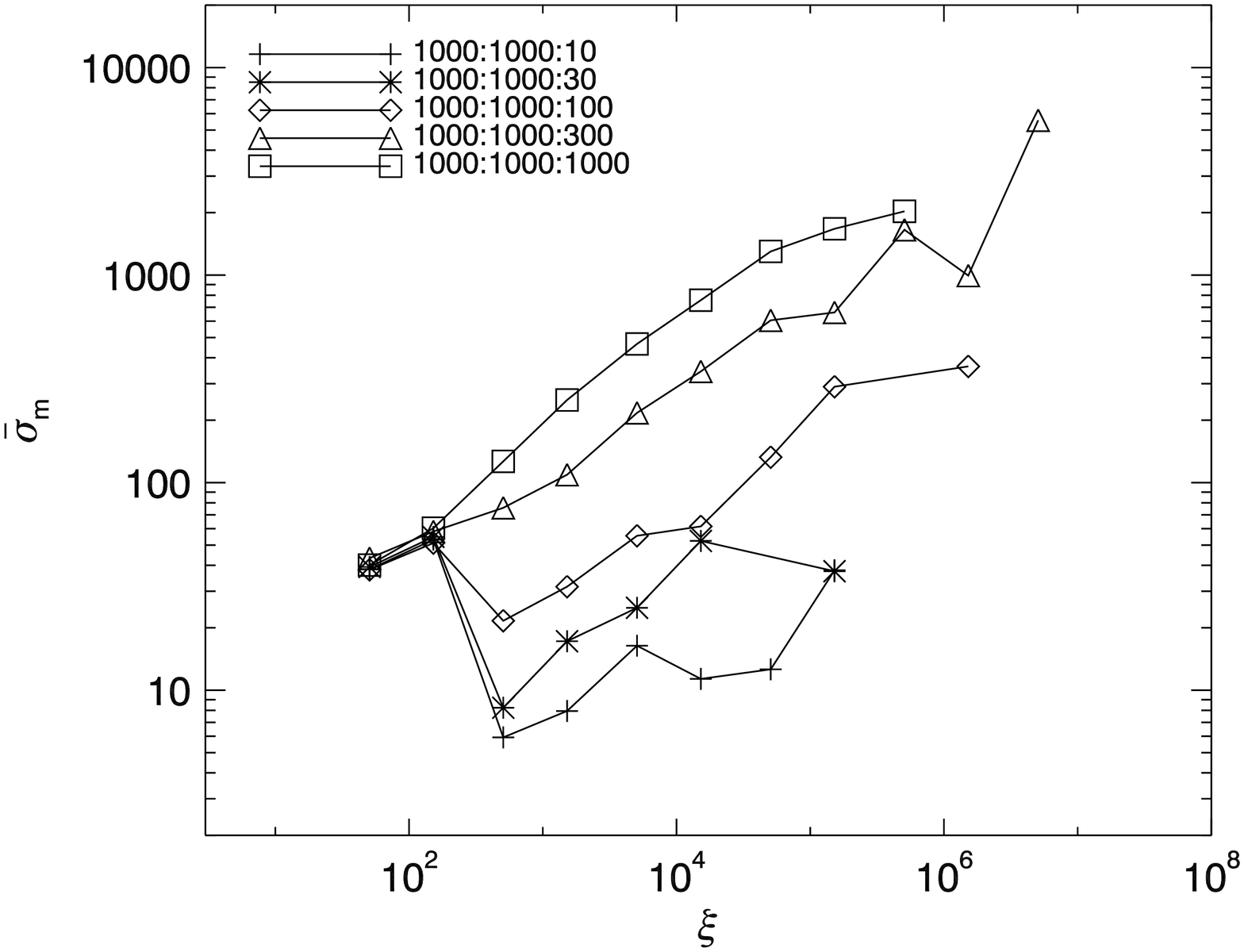}}
\caption{Normalized merger cross sections like
  Fig.~\ref{mergercs1010x} for 1000:1000:X series.  The error bars
  from Figure~\ref{mergercs1010x} are representative for the curves in
  this figure.}
\label{mergercs10001000x}
\end{figure}

\begin{figure}
\epsscale{0.75}
\rotatebox{90}{\plotone{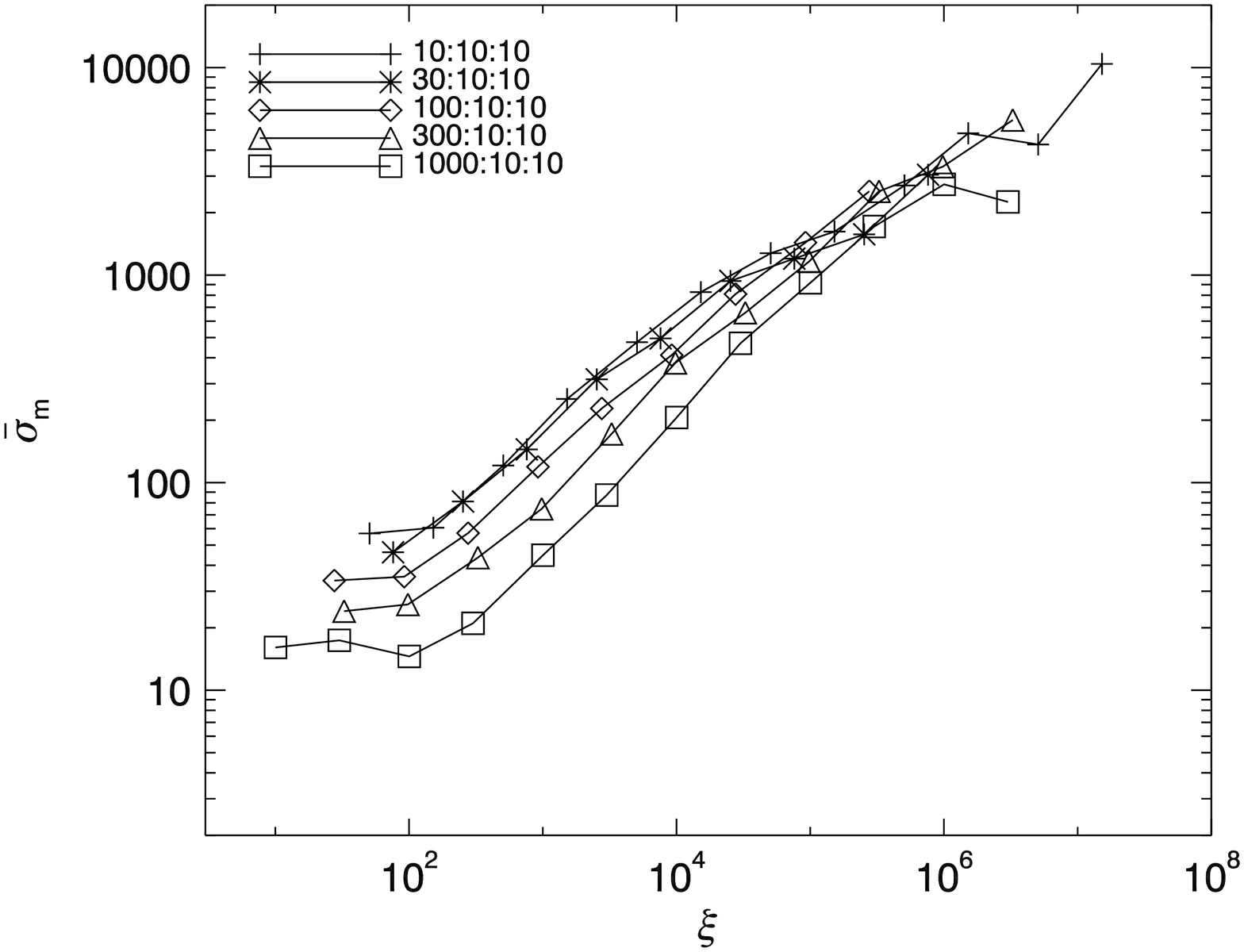}}
\caption{Normalized merger cross section like Fig.~\ref{mergercs1010x}
  for X:10:10 series.  The error bars from Figure~\ref{mergercs1010x}
  are representative for the curves in this figure.}
\label{mergercsx1010}
\end{figure}

\begin{figure}
\epsscale{0.75}
\rotatebox{90}{\plotone{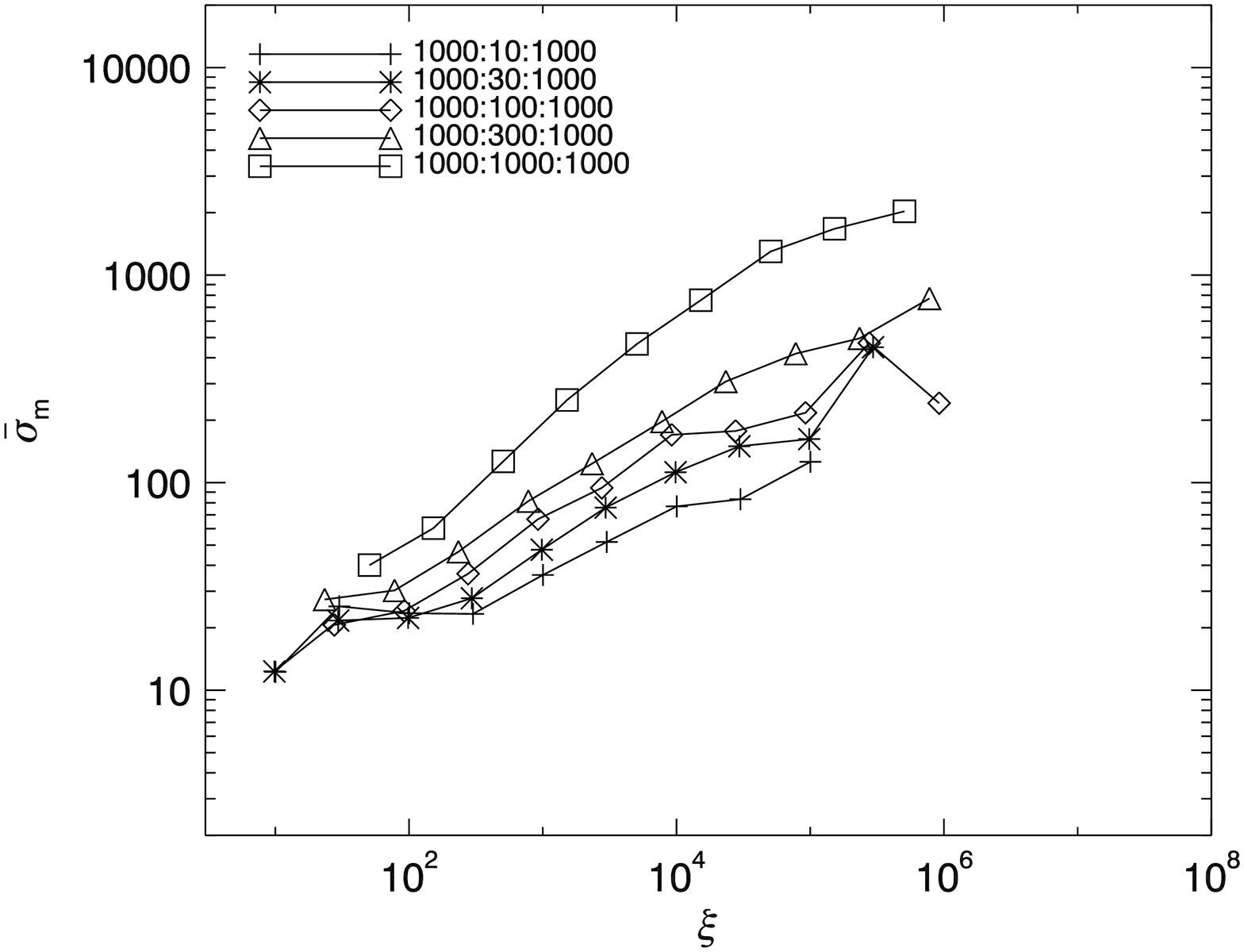}}
\caption{Normalized merger cross section like Fig.~\ref{mergercs1010x}
  for 1000:X:1000 series.  The error bars from
  Figure~\ref{mergercs1010x} are representative for the curves in this
  figure.}
\label{mergercs1000x1000}
\end{figure}

The most interesting new consequence from adding the effects of
gravitational radiation to the three-body problem is the possibility
of a merger between two objects.  Though the two-body cross section
for merger can be calculated from Equation~\ref{maxpericapture}, the
dynamics of three-body systems increases this cross section in a
nontrivial manner.  We present simulations of individual binary-single
encounters for a variety of masses.  As in Paper~I, the interactions
were set up in hyperbolic encounters with a relative velocity at
infinity of $v_{\infty} = 10\kms$ with an impact parameter
distribution such that $P(b)\propto b$ with $b_{\rm min} = 0$ and
$b_{\rm max}$ such that the maximum pericenter separation would be
$r_{p} = 5a$.  The binaries were initially circular with semimajor
axes ranging from $10^{-6}\au$ to $10^{2}\au$, depending on the mass.
The masses were picked such that one of the three mass ratios was
unity with all masses ranging from $10~\msun$ to $10^{3}\msun$ with
roughly half-logarithmic steps.  For each mass and semimajor axis
combination, we run $10^{4}$ encounters.  We calculate the merger
cross section as $\sigma_{m} = f \pi b^{2}_{\rm max}$ where $f$ is the
fraction of encounters that resulted in a merger while all three
objects were interacting.  In
Figures~\ref{mergercs1010x}-\ref{mergercs1000x1000} we plot, as a
function of the semimajor axis scaled to the gravitational radius of
the binary $\xi \equiv a / (G m_{\rm bin} / c^{2})$, the cross section
normalized to the physical cross section of the Schwarzschild radius
of the mass of the entire system taking gravitational focusing into
account:
\begin{equation}
\bar{\sigma}_{m} = {\sigma_{m}}{\left[4\pi \frac{GM_{\rm tot} }{ v^{2}_{\infty}}
\frac{GM_{\rm tot} }{ c^{2}}\right]^{-1}}. 
\label{sigmabar}
\end{equation}
For all mass ratios $\bar{\sigma}_{m}$ increases
with $\xi$ because hard binaries with wide separations sweep out
larger targets where the interloper can interact with and merge with
the binary components.  As $\xi$ increases to the point that the
binary is no longer hard, $\bar{\sigma}_{m}$ will approach the value
expected from Equation~\ref{maxpericapture}.  The curves flatten out
for $\xi \la 100$ as the cross section is dominated by the mergers of
binary members with each other because of hardening interactions and
eccentricity kicks that bring the two masses together.  For
sufficiently small $\xi$, the merger cross section would be formally
infinite since all binaries would merge quickly.  For all mass series,
as the mass ratios approach unity, the cross section increases
because complicated resonant encounters, which produce more numerous and
smaller close approaches, are more likely when all three objects are
equally important dynamically.

We note some interesting trends that can be seen in the plots.  Note
that for the scalings given, it is only the mass ratios that matter
and not the absolute mass so that the 10:10:10 and 1000:1000:1000
cases only differ because of statistical fluctuations
(Figs.~\ref{mergercs1010x} and~\ref{mergercs10001000x}).  Thus our
results can be scaled to others, e.g., 1000:100:100 would be the
same as 100:10:10.  For the 10:10:X mass series
(Fig.~\ref{mergercs1010x}), the normalized cross section decreases
with increasing interloper mass, roughly as $\bar{\sigma}_{m} \sim
(m_{2} / m_{\rm bin})^{-1}$. This happens because as the interloper
dominates the total mass of the system, complicated resonant
interactions with more chances for close approach are less likely.
Thus for the 10:10:1000 case, there are far fewer chances for a close
approach that results in merger.  The 1000:1000:X series (Fig.~\ref{mergercs10001000x}) shows a
distinct break around $\xi \sim 100$.  Since the binary mass is the
same for all curves, they all approach the same value for $\xi \la
100$ where the binary members merge with each other because of their
small separation.  For $\xi \ga 100$, the higher mass interlopers are
dynamically more important and cause more mergers.  The X:10:10 series
curves (Fig.~\ref{mergercsx1010}) all approach the 10:10:10 curve for $\xi \ga 10^{5}$ where the
dominant object in the binary has less influence over its companion.

\subsection{Sequences of Encounters}
\label{sequences}

Because a tight binary in a dense stellar environment will suffer
repeated encounters until it merges from gravitational radiation, we
simulate a binary undergoing repeated interactions through sequences
of encounters including gravitational radiation reaction.  As in
Paper~I, we start with a circular binary with initial semimajor axis
$a_{0} = 10\au$ and a primary of mass $m_{0} =$ 10, 20, 30, 50, 100,
200, 300, 500, or 1000~$\msun$ and a secondary of mass $m_{1} =
10~\msun$.  We simulate encounters with interloping black holes with
mass $m_{2} = 10~\msun$.  After each encounter, we integrate
Equations~\ref{petersa} and~\ref{peterse} to get the initial semimajor
axis and eccentricity for the next encounter.  This procedure
continues until the binary merges from gravitational radiation or
there is a merger during the encounter.  Throughout our simulations we
use an encounter velocity of $v_{\infty} = 10~\mathrm{~km~s^{-1}}$, an
isotropic impact parameter such that the hyperbolic pericenter would
range from $r_{p} = 0$ to $5a$, and a black hole number density in the
core $n = 10^{5}~\mathrm{pc}^{-3}$ (See Paper~I for an explanation of
these choices.).  For each mass ratio we simulate 1000 sequences of
encounters with gravitational radiation reaction. 

Our results are summarized in Table~\ref{coreresults}. The inclusion
of gravitational waves during the encounter makes a significant
difference from the results reported in Paper~I.  The fraction of
sequences that result in a merger during an encounter $f_{m}$ is a
good indicator of the importance of gravitational waves.  Even for
$m_{0} = 10~\msun$, a significant fraction ($f_{m} > 0.1$) of the
sequences merge this way, and for $m_{0} > 300~\msun$ this type of
merger is more likely to occur than mergers between encounters, and
thus this effect shortens the sequence significantly.  In particular,
for $m_{0} = 1000~\msun$ compared to the values from Paper~I, the
average number of encounters per sequence
$\left<n_{\mathrm{enc}}\right>$ is decreased by 42\%; the average
number of black holes ejected from the cluster
$\left<n_{\mathrm{ej}}\right>$ is reduced by 56\%; and the average
sequence length $\left<t_{\mathrm{seq}}\right>$ is 67\% shorter.  One
caveat for the study of sequences of encounters is that an IMBH in a
cluster of much lower mass objects will gather a large number of
companions in elongated orbits through binary disruptions, and thus
the picture of an isolated binary encountering individual black holes
may not hold when the IMBH becomes very massive \citep{pfahl05a}.

\begin{deluxetable}{rrrrrrrr}
  \footnotesize
  \tablecaption{Sequence Statistics}
  \tablehead{
    \colhead{$m_{0}/\msun$} &
    \colhead{$\left<n_{\mathrm{enc}}\right>$} &
    \colhead{$\left<n_{\mathrm{ej}}\right>$} &
    \colhead{$f_{\mathrm{binej}}$} &
    \colhead{$\left<t_{\mathrm{seq}}\right>/10^{6}~\mathrm{yr}$} &
    \colhead{$\left<a_{f}\right>/\mathrm{AU}$} &
    \colhead{$\left<e_{f}\right>$} &
    \colhead{$f_{m}$}
  }
  \startdata
  10   &  46.4 &   3.2 &  0.652 & 54.10 &  0.174 & 0.904 & 0.134 \\
  20   &  46.7 &   5.1 &  0.515 & 40.86 &  0.224 & 0.900 & 0.130 \\
  30   &  52.4 &   7.3 &  0.457 & 29.47 &  0.290 & 0.898 & 0.156 \\
  50   &  62.3 &  10.8 &  0.329 & 19.17 &  0.291 & 0.897 & 0.190 \\
 100   &  83.9 &  16.6 &  0.103 & 11.65 &  0.401 & 0.893 & 0.275 \\
 200   & 123.0 &  24.3 &  0.011 &  7.26 &  0.411 & 0.885 & 0.387 \\
 300   & 147.8 &  26.9 &  0.002 &  4.74 &  0.543 & 0.881 & 0.492 \\
 500   & 197.5 &  33.1 &    -   &  3.03 &  0.611 & 0.879 & 0.627 \\
1000   & 284.2 &  38.8 &    -   &  1.47 &  0.878 & 0.914 & 0.754
  \enddata
  \label{coreresults}
  \tablecomments{Main results of simulations of sequences of
    encounters with gravitational radiation included during the
    encounter.  The columns are: the mass of the dominant black hole
    $m_{0}$, the average number of encounters per sequence
    $\left<n_{\mathrm{enc}}\right>$, the average number per sequence
    of stellar-mass black holes ejected from a stellar cluster with
    escape velocity $50\kms$ $\left<n_{\mathrm{ej}}\right>$, the
    fraction of sequences in which the binary is ejected
    $f_{\mathrm{binej}}$ from a stellar cluster with escape velocity
    $50\kms$, the average time per sequence
    $\left<t_{\mathrm{seq}}\right>$, the average final semimajor axis
    of the binaries after the last encounter $\left<a_{f}\right>$, the
    average final eccentricity of the binaries after the last
    encounter $\left<e_{f}\right>$, and the fraction of sequences that
    end with a merger during the encounter $f_{m}$.  Note that
    $\left<a_{f}\right>$ and $\left<e_{f}\right>$ only refer to the
    binaries that do not merge during the encounter; these comprise $1
    - f_{m}$ of the sequences.  }
\end{deluxetable}

\section{Discussion}
\label{discussion}

\subsection{Implications for IMBH Formation and Growth}
\label{imbhformation}

Our simulations provide a useful look into the merger history of an
IMBH or its progenitor in a dense stellar cluster.  As an IMBH grows
through mergers with stellar-mass black holes, it will progress
through the different masses that we included in our simulations of
sequences.  We interpolate the results in Table~\ref{coreresults} to
calculate the time it takes to reach $1000~\msun$, the number of
cluster black holes ejected while building up to $1000~\msun$, and the
probability of retaining the IMBH progenitor in the cluster for
different seed masses and escape velocities of the cluster.

\begin{figure}
\epsscale{0.75}
\rotatebox{90}{\plotone{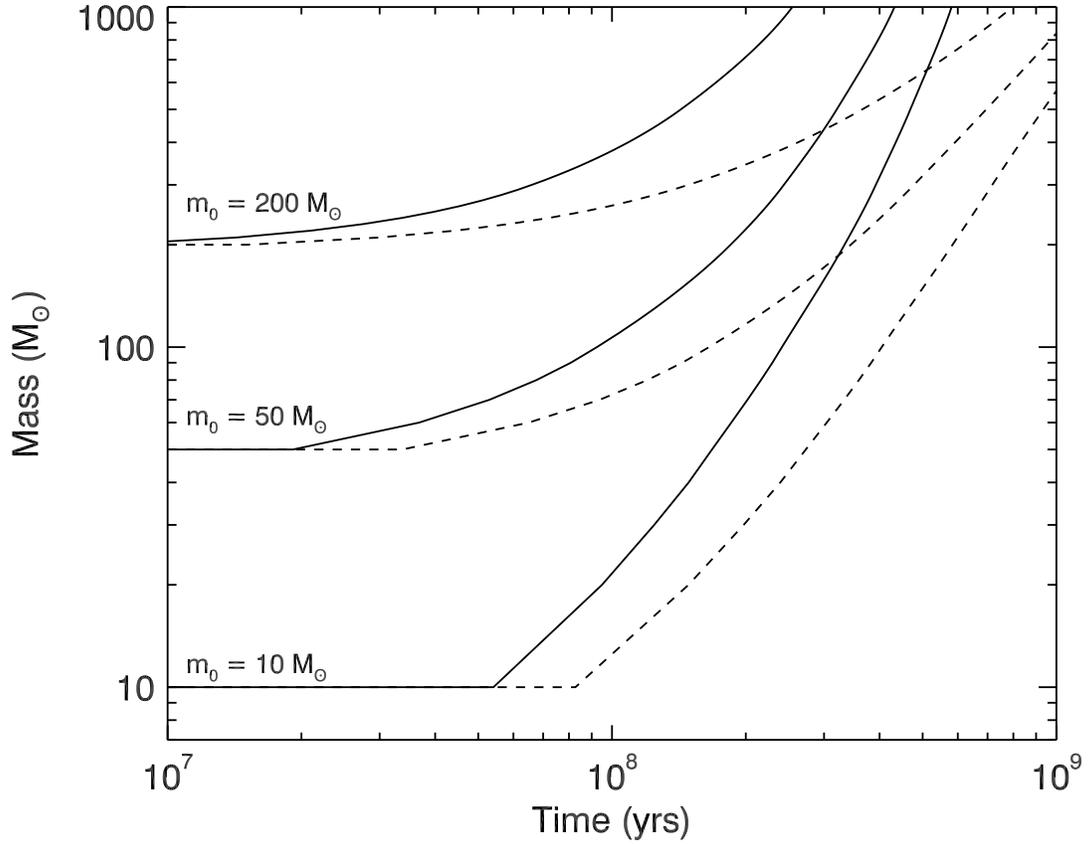}}
\caption{Mass of progenitor IMBH as a function of time as it grows
  through mergers with $10~\msun$ black holes in a dense stellar
  cluster.  Solid curves show results from this work in which
  gravitational radiation is included, and dashed curves show results
  from Paper~I in which this effect is ignored.  From bottom to top
  the curves show the growth of black holes with initial mass $m_{0} =
  10$, $50$, and $200~\msun$.  The IMBH progenitors all reach
  $1000~\msun$ in less than $600\units{Myr}$, and the inclusion of
  gravitational radiation significantly speeds up the growth of the
  black hole.}
\label{imbhtime}
\end{figure}

The time to build up to $1000~\msun$ is dominated by
$\left<t_{\mathrm{seq}}\right>$ at high masses.  Although each
individual sequence is short, far more mergers are required for the
same fractional growth in mass.  In Figure~\ref{imbhtime} we plot the
mass of the IMBH as a function of time for an initial mass of $m_{0} =
10$, $50$, and $200~\msun$, for which total times to reach
$1000~\msun$ are $600$, $400$, and $250~\units{Myr}$, respectively.
Because we assume a constant core density throughout the simulations,
the times are unaffected by changing the cluster's escape velocity.
Without gravitational radiation, the times are roughly twice as long
(Paper~I) because the length of each sequence is dominated by the time
it spends between encounters at small $a$ when encounters are rarer.
With gravitational radiation included, mergers that occur during an
encounter are more likely at small separations, and the length of the
sequence is shortened.  These times are much shorter than the age of
the globular cluster and are smaller than or comparable to timescales
for ejection of black holes from the cluster, which we discuss below
\citep[see also][]{pzm00, olearyetal05}.  Thus time is not a limiting factor in
reaching $1000~\msun$ for an IMBH progenitor that can remain in a
dense cluster with a sufficiently large population of stellar mass
black holes.

\begin{figure}
\epsscale{0.75}
\rotatebox{90}{\plotone{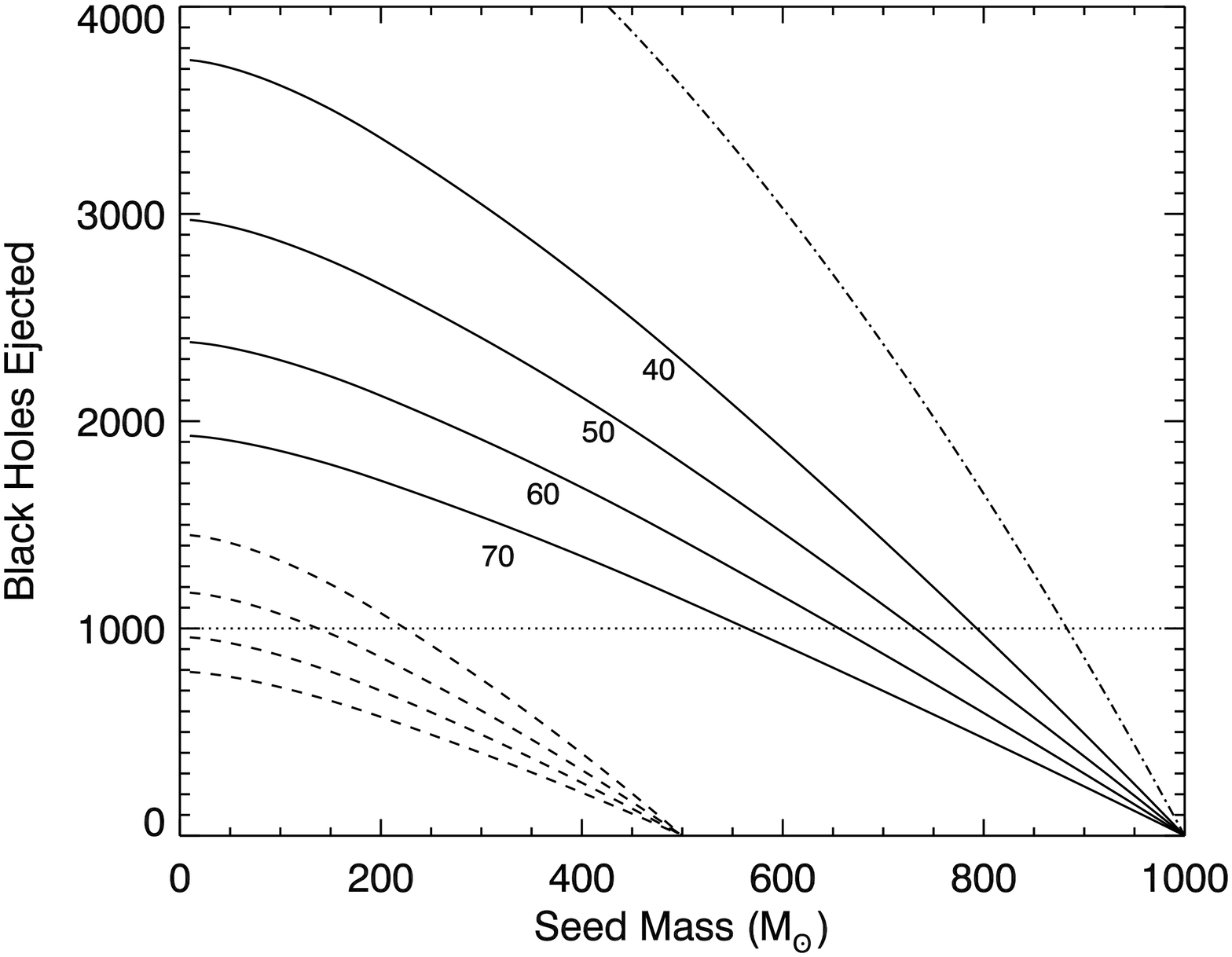}}
\caption{Number of black holes ejected in building up to $1000~\msun$
  (solid curves) and to $500~\msun$ (dashed curves) as a function of
  seed mass for different cluster core escape velocities, given in
  units of $\mathrm{km~s^{-1}}$.  The dotted line indicates the
  expected number of black holes in a dense globular cluster.  The
  dot-dashed curve from Paper~I shows the number of black holes
  ejected from the cluster in building up to $1000~\msun$ for a
  cluster escape velocity of $50\kms$ without the effects of
  gravitational waves during the encounter.  For all but the largest
  seed masses, the globular cluster does not contain enough black
  holes for the IMBH to reach $1000~\msun$.  There are, however, a
  sufficient number of black holes to build up to $500~\msun$ for a
  seed mass greater than $225~\msun$ or an escape velocity of at least
  $60\kms$.  The inclusion of gravitational radiation during the
  encounter roughly halves the number of ejections.}
\label{imbhfieldejections}
\end{figure}

Each time that an encounter tightens the binary, energy is transfered
to the interloper, which leaves with a higher velocity.  If energetic
enough, this interaction will kick the interloper out of the cluster.
If the interactions kick all of the interacting black holes out of the
cluster, the IMBH cannot continue to grow.  In a dense cluster, there
are roughly $10^{3}$ black holes (Paper~I).  With gravitational
radiation included during the encounters, the number of black holes
ejected is roughly halved (Fig.~\ref{imbhfieldejections}), but the
total number ejected while building up to $1000~\msun$ is still a few
times the number of black holes available even for an escape velocity
of $v_{\rm esc} = 70\kms$.  Thus a black hole smaller than $m_{0} \la
600~\msun$ cannot reach $1000~\msun$ by this method without additional
processes such as Kozai resonances \citep[though
\citealt{olearyetal05} find that Kozai-resonance induced mergers
will only increase the total number of mergers by
$\sim10\%$]{gmh04,mh02b,wen03}.  There is still the potential for
significant growth in a short period of time.  If we consider the
point at which half of the black holes have been ejected from the
cluster as the end of growth, then a black hole with initial mass of
$50~\msun$ will grow to $290~\msun$ in $120~\units{Myr}$, and a black
hole of $200~\msun$ will grow to $390~\msun$ in less than
$100~\units{Myr}$.  In addition, this ejection of stellar-mass black
holes by a binary with a large black hole is faster than by
self-ejection from interactions among stellar-mass black holes
calculated by \citet{pzm00}, who find that $\sim90\%$ of black holes
are ejected in a few Gyr. \citet{olearyetal05}, however, find that the
inclusion of a mass spectrum of black holes further speeds up the
ejection of stellar-mass black holes.

\begin{figure}
\epsscale{0.75}
\rotatebox{90}{\plotone{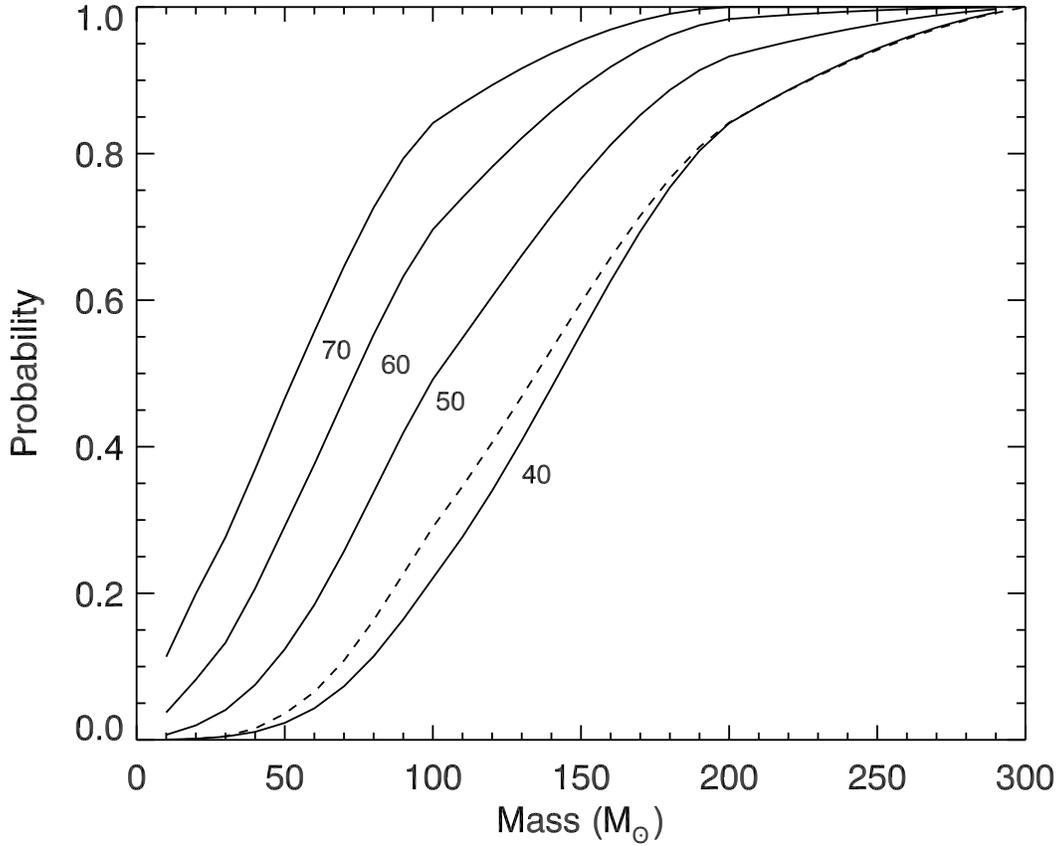}}
\caption{Probability for a binary with an IMBH to remain in the
  cluster until building up to 300~\msun\ as a function of seed mass
  for different cluster core escape velocities given in units of
  $\mathrm{km s^{-1}}$.  Solid curves are results from this work, and
  the dashed curve is from Paper~I for an escape velocity of $50\kms$.
  The inclusion of gravitational radiation significantly increases the
  retention probability.}
\label{imbhbinaryejections}
\end{figure}

For every kick imparted on an interloper, conservation of momentum
ensures a kick on the binary.  Even with a large black hole, extremely
large kicks can eject the binary from the cluster, at which point the
IMBH progenitor can no longer grow.  We can calculate the probability
of IMBH retention for an individual sequence as $1 -
f_{\mathrm{binej}}$, from which we interpolate the probability of
remaining in the cluster while growing to $300~\msun$ when the binary
is essentially guaranteed to remain in the cluster.  We plot this
probability as a function of seed mass for several different escape
velocities in Figure~\ref{imbhbinaryejections}.  The inclusion of
gravitational waves during the encounter increases the retention
probability for small masses.  For $m_{0} = 50~\msun$ the cluster
retains the binary more than 12\% of the time, and 49\% of the time
for $m_{0} = 100~\msun$.  Because the energy that an interloper can
carry away from the system scales as
\begin{equation} 
\Delta E \sim \frac{m_{1}}{m_{0} + m_{1}} \left|E_{B}\right| =
\frac{m_{1}}{m_{0} + m_{1}} \frac{Gm_{0}m_{1}}{2a},
\end{equation} 
the encounters at the end of the sequence, when $a$ is smallest, are
the most likely to impart a kick large enough to eject the binary from
the cluster.  This is also the point at which effects from
gravitational radiation are strongest and at which close encounters
are most likely to cause a merger.  When the encounter ends in a
merger, there can be no more ejections.  The mergers from
gravitational radiation decrease the number of ejections by decreasing
the number of encounters and thus the number of possible ejections as
well as cutting off what would otherwise be the end of the sequence,
in which ejections are more likely to occur.

Our analysis of the ejection of stellar-mass black holes as well as of
IMBH progenitors does not include the effects of gravitational
radiation recoil.  As two objects with unequal masses or with
misaligned spins spiral in towards each other, asymmetric emission of
gravitational radiation produces a recoil velocity on the center of
mass of the binary.  Most of the recoil comes from contributions after
the masses are inside of the ISCO, where post-Newtonian analysis
becomes difficult \citep{fhh04}.  For non-spinning black holes, the
velocity kick from the recoil up to the ISCO is \citep{fhh04}
\begin{equation}
v_{r} = 15.6\kms \frac{f(q)}{f_{\rm max}}
\label{iscorecoil}
\end{equation}
where $q = m_{1} / m_{0} < 1$, $f(q) = q^{2}(1 - q)/(1 + q)^{5}$, and
$f_{\rm max} \approx f(0.38) = .018$.  \citet{fhh04} bounded the total
recoil to between $20\kms \le v_{r} \le 200\kms$ for non-spinning
black holes with $q=0.127$.  Since the recoil velocity scales as
$q^{2}$ for $q \ll 1$, this may be scaled to other mass ratios.  More
recently, \citet{bqw05} argued from high order post-Newtonian
expansions that the kick speed for very small mass ratios $q\ll 1$ was
$v_{r}/c=0.043q^2$, with an uncertainty of roughly 20\%.  This is
consistent with the results of \citet{fhh04}, but as most of the
recoil originates well inside the ISCO, \citet{bqw05} caution that
numerical results may be required for definitive answers.  In both
cases, a seed mass of $m_{0} = 150~\msun$ merging with $m_{1} =
10~\msun$ companions will produce a recoil velocity $v_{r} \la
50\kms$.  A seed mass greater than 150~\msun will then avoid ejection
both from dynamical interactions and from gravitational radiation
recoil.

\subsection{Implications for Gravitational Wave Detection}
\label{gwdetection}

In addition to the likelihoods and rates of growth of black holes in
dense stellar systems, our simulations shed light on the gravitational
wave signals that come from the mergers of these black holes.  Making
optimistic assumptions, \citet{olearyetal05} calculate upper limits
for Advanced LIGO detection rates of all black hole mergers in stellar
clusters formed at a redshift $z = 7.8$.  For their wide range of
cluster properties, they find detection rates ranging from $\nu_{\rm
  LIGO} \approx 0.6$ to $10\units{yr^{-1}}$.  For cluster parameters
that most closely resemble those used in Paper~I and in this work (GMH
model series), they find $\nu_{\rm LIGO} \approx 2$ to
$4\units{yr^{-1}}$.  Our simulations show that when gravitational
radiation is included in the integration the number of black holes
ejected per merger decreases for all mass ratios.  With fewer black
holes ejected from the cluster, the overall rate of black hole mergers
increases.  For the 10:10:10 case, the number of ejections per merger
decreases by $\sim10\%$, and for the 1000:10:10 case the number
decreases by more than a factor of $2$, thus increasing the rates
found by \citet{olearyetal05}.  The exact increase in rate is
difficult to estimate because the total number of mergers is dominated
by mergers between stellar-mass black holes yet the most easily
detected mergers involve black holes with larger masses.

Because dynamical interactions strongly affect the eccentricity of a
binary and because the timescale for merger is a such a strong
function of eccentricity, binaries in a cluster tend to have very high
eccentricities after their last encounter \citep[Paper~I,
][]{olearyetal05}.  With the addition of gravitational radiation
during the encounter, we find that the merging binaries become more
eccentric because a significant fraction of the mergers ($f_{m}$ in
Table~\ref{coreresults}) occur during the encounter.  These mergers
typically happen between two black holes that are not bound to each
other until they come close to each other and emit a significant
amount of gravitational radiation, after which the two black holes are
in an extremely high eccentricity orbit ($1 - e \la 10^{-3}$).

\begin{figure}
\epsscale{0.75}
\rotatebox{90}{\plotone{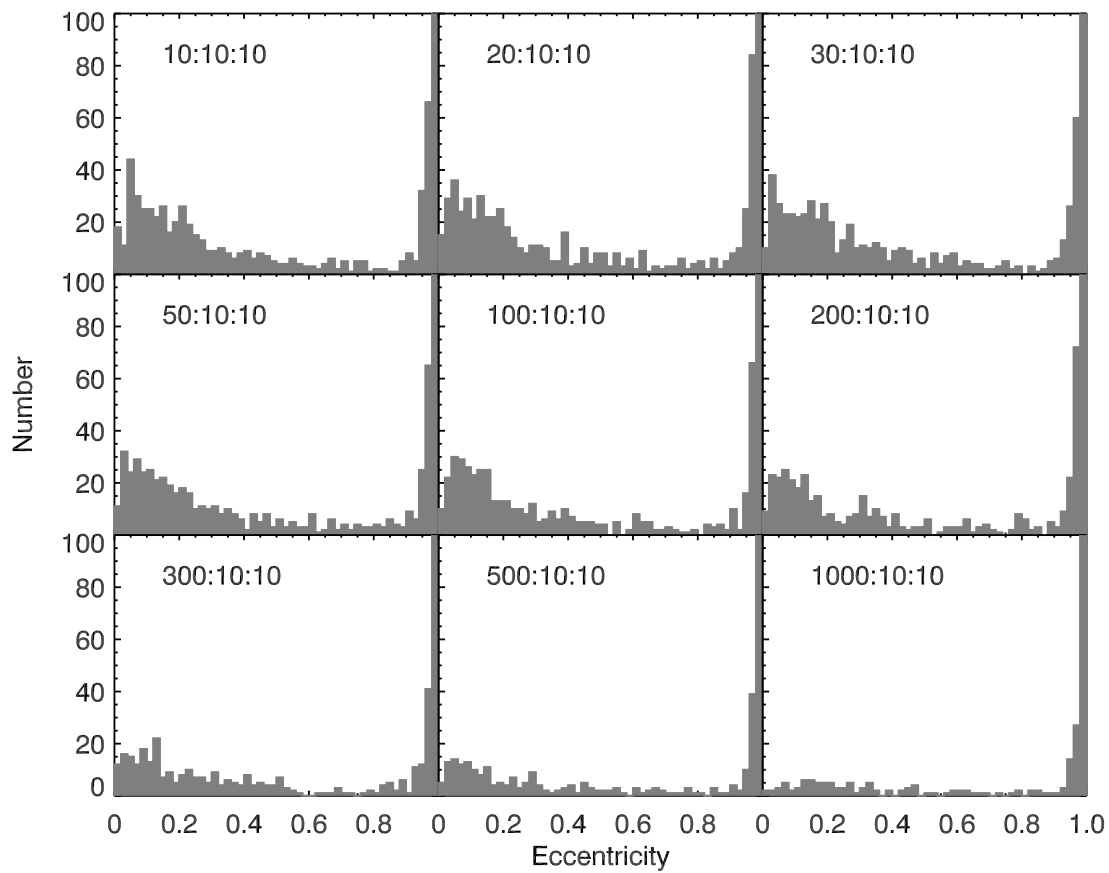}}
\caption{Histogram of eccentricities of merging binary while in the
  \emph{LISA} band ($f_{GW} = 2\mhz$ to $10\mhz$) out of a total of
  1000 sequences.  The histograms show a combination of the binaries
  that merged after the last encounter with eccentricities
  concentrated around $0 < e \la 0.3$ and the black holes that merged
  quickly during the encounter with eccentricities very close to
  unity.  The peaks in the rightmost bin in all plots lie above the
  range of the plots.}
\label{lisaecc}
\end{figure}

\begin{figure}
\epsscale{0.75}
\rotatebox{90}{\plotone{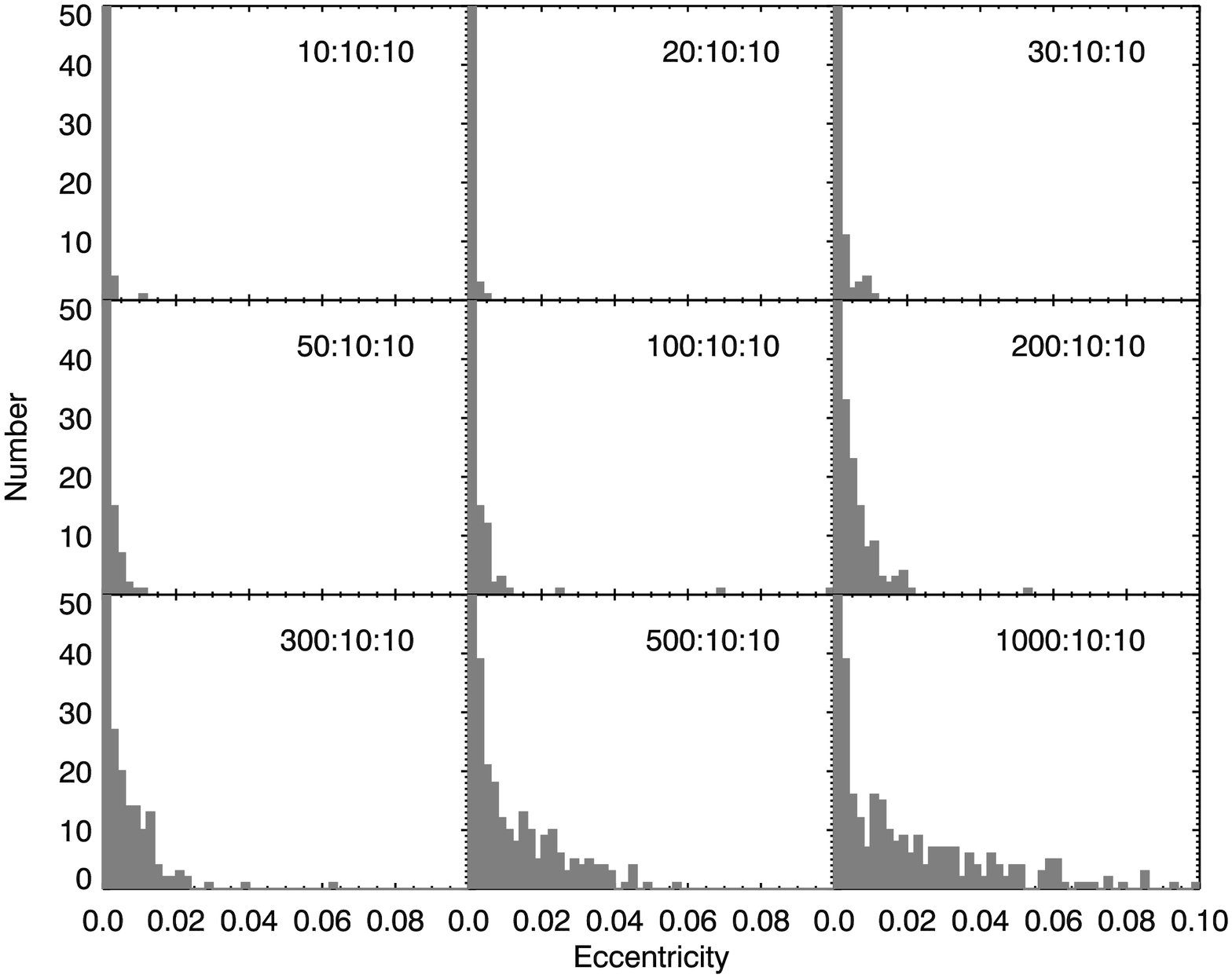}}
\caption{Histogram of eccentricities of merging binary while the
  gravitational wave frequency is detectable from current and future
  ground-based detectors.  The upper limit of the frequency range is
  the ISCO frequency.  We used a lower limit for frequency range of
  $100\hz$ for $m_{0} = 10$, and $20~\msun$; $35\hz$ for $m_{0} = 30$,
  $50$, and $100~\msun$; and half the ISCO frequency for the higher
  mass binaries.  The binaries are very close to circular once they
  are in the frequency range of ground-based detectors.The peaks in
  the rightmost bin in all plots lie above the range of the plots.}
\label{ligoecc}
\end{figure}

To see how these high eccentricities affect the detectability of the
gravitational wave signal, we integrate Equations~\ref{petersa} and
\ref{peterse} until the binaries are detectable by \emph{LISA} and
then Advanced LIGO.  For circular orbits, the frequency of
gravitational wave emission is twice the orbital frequency, but
masses in eccentric orbits emit at all harmonics: $f_{\rm GW}= n
\Omega / 2\pi$, where $n$ is the harmonic number and
\begin{equation}
\Omega = \left[\frac{G\left(m_{0} + m_{1}\right)}{a^{3}}\right]^{1/2}
\label{eccfreq}
\end{equation}
with peak harmonic for $e > 0.5$ at approximately $n = 2.16\left(1 -
  e\right)^{-3/2}$ \citep{fp03}.  We consider the binary to be
detectable by \emph{LISA} when the peak harmonic frequency is between
$2\units{mHz}$ and $10\units{mHz}$.  We plot the distribution of
eccentricities in Figure~\ref{lisaecc}.  The distributions are
essentially a combination of those from Figure~9 of Paper~I and a
sharp peak near $e = 1$, which comes from the mergers during the
encounter.  The number in the sharp peak increases with mass as
$f_{m}$ increases such that for 1000:10:10 more than 75\% of the
merging binaries detectable by LISA have an eccentricity greater than
0.9.  Between 15\% and 25\% of all of the merging binaries have
eccentricities so high that the peak harmonic frequency is above the
most sensitive region of the LISA band, but they should still be
emitting strongly enough at other harmonics to be detectable.  Such
high eccentricity presents challenges for the detection of these
signals from the data of space-based gravitational wave detectors
because (1) it requires a more computationally expensive template
matching that includes non-circular binaries and (2) the binaries only
emit a strong amount of gravitational radiation during the short time
near periapse as they merge.  For a given semimajor axis, these
extremely high eccentricities will also increase the gravitational
wave flux emitted and thus increase the distance out to which
\emph{LISA} can detect them, but the detection rate may be compensated
by the fact that more parameters are required \citep{will04}.  We also
integrate the orbital elements of the binaries until they are in the
Advanced LIGO band ($40\units{Hz} < f_{\rm GW} < f_{\rm ISCO}$) or
within a factor of 2 of their ISCO frequency for $m_{0} > 100~\msun$.
We find that they have almost completely circularized
(Fig.~\ref{ligoecc}).  A tiny fraction ($<0.5\%$) of the runs
with $m_{0} = 500$ and $1000~\msun$ have merging binaries with
eccentricities such that $1 - e \la 10^{-6}$.

\section{Conclusions}
\label{conclusions}
\emph{1. Gravitational radiation in \emph{N}-body.}  We present
results of numerical simulations of binary-single scattering events
including the effects of gravitational radiation during the encounter.
We include gravitational radiation by adding the 2.5-order
post-Newtonian force term (Eq.~\ref{hndgwdrag}) to the equation of
motion within the HNDrag framework.  The code reproduces the expected
semimajor axis and eccentricity evolution, and it gives the expected
two-body capture radius.

\emph{2. Close approach and merger cross sections.}  We use the new
code to test the effects of gravitational radiation on a standard
numerical experiment of binary-single encounters.  We probe the close
approach cross section to smaller separations than has been simulated
previously and find that the inclusion of gravitational radiation
makes little difference except for extremely close encounters ($r_{p}
< 10^{-5}a$), at which point gravitational radiation drives the
objects closer together.  We also present the cross section for merger
during binary-single scattering events for a variety of mass ratios
and semimajor axes.

\emph{3. IMBH growth in dense stellar clusters.}  We simulate
sequences of binary-single black hole encounters to test for the
effects of gravitational radiation and to test formation and growth
models for intermediate-mass black holes in stellar clusters.  We find
that the inclusion of gravitational radiation speeds up the growth of
black holes by a factor of 2, increases the retention of IMBH
progenitors by a factor of 2, and decreases the ejection of
stellar-mass black holes by a factor of 2.  All of these effects act
to enhance the prospects for IMBH growth.

\emph{4. Detectability of gravitational waves.}  We analyzed the
merging binaries from the simulations of black holes in dense stellar
clusters to look at the detectability of the gravitational wave
signals from these sources.  We find that the mergers that occur
rapidly during the encounter as opposed to those that occur after the
final encounter are an important source of black hole mergers,
becoming the dominant source of mergers at the higher mass ratios.
The mergers that do occur during the encounter tend to have extremely
high eccentricity ($e > 0.9$) while in the \emph{LISA} band,
presenting challenges for their detection.  When the gravitational
wave signal from the merging black holes is in the Advanced LIGO band,
the orbit will have completely circularized.

\acknowledgements We are grateful for the hospitality of the Center
for Gravitational Wave Physics and for suggestions from the anonymous
referee.  Many of the results in this paper were obtained using
VAMPIRE, the Very Awesome Multiple Processor Integrated Research
Environment, and the Beowulf cluster of the University of Maryland
Department of Astronomy.  This research has made use of NASA's
Astrophysics Data System.  This work was supported by NASA grant NAG
5-13229.

\bibliographystyle{astroads}
\bibliography{gultekin}

\end{document}